\documentclass[aps,pra,reprint,groupedaddress]{revtex4-1}
\usepackage{amsmath,graphicx,hyperref}
\usepackage[FIGTOPCAP,raggedright,nooneline,bf,footnotesize]{subfigure}
\usepackage{indentfirst}
\usepackage[usenames,dvipsnames]{pstricks}

\newcommand{\figref}[1]{Fig.~\ref{#1}}
\renewcommand{\eqref}[1]{Eq.~\ref{#1}}
\newcommand{\appref}[1]{Appendix~\ref{#1}}

\newcommand{\prop}{\mathcal{S}}
\usepackage{overpic}
\usepackage{xcolor}
\usepackage{pict2e}

\begin{document}

%\title{Chromatic dispersion effects on spectrally correlated photon pairs propagating in a single mode fiber}

%\title{Propagation of a spectrally correlated photon pair in a dispersive single mode fiber}
%\title{Increasing the range of quantum communication using spectral entanglement}
%\title{Pushing up the limit of quantum communication distance using spectral entanglement}
\title{Reducing detection noise of a photon pair in a dispersive medium by controlling its spectral entanglement}

\author{Karolina Sedziak}
\author{Miko\l aj Lasota}
\author{Piotr Kolenderski}
\email{kolenderski@fizyka.umk.pl}
\affiliation{Faculty of Physics, Astronomy and Informatics, Nicolaus Copernicus University, Grudziadzka 5, 87-100 Toru\'{n}, Poland} 

\date{\today}

\begin{abstract}
Chromatic dispersion is one of the main limitations to the security of quantum communication protocols that rely on the transmission of single photons in single mode fibers. This phenomenon forces the trusted parties to define longer detection windows to avoid losing signal photons and increases the amount of detection noise that is being registered. In this work, we analyze the effects of chromatic dispersion on a photon pair generated via spontaneous parametric down-conversion and propagating in standard telecommunication fibers. We also present the possibility of reducing the detection noise by manipulating the spectral correlation of the pair. As an example, we show that our results can be used to increase the maximal security distance of a discrete-variable quantum key distribution scheme in which the photon source is located between the legitimate participants of the protocol.
\end{abstract}

\pacs{42.50.Dv, 03.67.Hk, 03.67.Dd}

\maketitle

\section{Introduction}
Single photon sources are essential for the experimental implementation of various quantum information processing and communication protocols. One of the most popular types is based on the process of spontaneous parametric down-conversion (SPDC) \cite{Louisell1961,Wasilewski2006,Shimizu2009,Bruno2014}. 
%Among their advantages one can certainly list photon production efficiency \cite{Pomarico12,DaCunha13,Ramelow13}, nonclassicality of the emitted photons \cite{Fasel04} and low cost in comparison to other similar devices.
However, the SPDC photons are typically broadband, which manifests in temporal broadening during their propagation in a dispersive medium \cite{Lutz2014}, such as single mode fibers (SMF). This effect has deteriorating influence on the performance of quantum communication protocols. This is because one of the most important factors limiting their security is the detection noise and its probability of occurrence is proportional to the duration of the photon detection window. Unfortunately, the chromatic dispersion effects due to the propagation in SMFs limit the possibility of narrowing the time window.

Nevertheless, the  effects can be reduced or even canceled \cite{Franson1992,Franson2009}. Local and nonlocal dispersion cancellation has been experimentally demonstrated \cite{Kaltenbaek2008,Lavoie2009a,Mazurek2013,Resch2007,Steinberg1992} and some of the resulting ideas were applied to measure the dispersion of the biphoton \cite{Barak2012}. The effects have also been analyzed in the context of realistic SPDC sources \cite{Perina1999}. Next, it was shown that the spectra of fiber-coupled SPDC photon pairs can be correlated negatively \cite{Wasilewski2006,Jin2014}, positively \cite{Kim2002,Kuzucu2008,Lutz2013,Lutz2014,Shimizu2009} or not correlated  at all \cite{Mosley2008,Sanchez-Lozano2012,Jin2013,Bruno2014}. While the possibility to enhance the security of quantum protocols by tailoring the photon pair state has been applied previously as a countermeasure against polarization mode dispersion \cite{Shtaif2011,Xiang2014}, to the best of our knowledge it has not been considered yet for reducing the effect of chromatic dispersion.

In this work, we theoretically analyze the propagation of a SPDC  pair through a pair of SMFs over distances long enough for chromatic dispersion effects to be pronounced. We investigate how the spectral correlation can influence the possibility to reduce the detection noise. Our results are subsequently applied to the security analysis of a discrete-variable quantum key distribution (DV QKD) scheme with a SPDC source of photons located in the middle between the legitimate participants, called Alice and Bob. We show that when the global time reference, i.e. the timing information of the pump laser pulses, is not available to the participants of the protocol, Alice and Bob can achieve considerably longer maximal security distance by utilizing the photon pairs featuring positive spectral correlation as compared to negative. This is possible as long as the photons are relatively broadband. On the other hand when the global time reference is available, the QKD security distance can be maximized by using strongly correlated (decorrelated) photon pairs if their spectra are broad (narrow). 
%In this case the sign of the spectral correlation does not matter. 
%\section{Results}
\section{Dispersion effects}

Let us assume that SPDC photons are coupled into a pair of SMFs of length $L$. The propagation is described by the  unitary transformation of the initial wave function, $\psi(t_1',t_2')$ \cite{Weiner2011}:
\begin{equation}
\psi_L(t_1,t_2)=\int\mathrm{d}t_1'\,\mathrm{d} t_2'\, \prop_1 (t_1,t_1',L) \prop_2(t_2,t_2',L)\psi(t_1',t_2'), 
\label{eq:propagation}
\end{equation} 
where $\psi_L(t_1,t_2)$ is the resulting state at the output of the fibers and  
\begin{equation}
\prop_k(t_k,t_k',L)=\frac{1}{\sqrt{4\pi i\beta L}}\exp\left(\frac{i(t_k-t_k')^2}{4 \beta L}\right)
\label{eq:propagator}
\end{equation}
is a propagator introducing the evolution. The chromatic dispersion is taken into account up to second order in terms of frequency detuning, $\nu=\omega-\omega_0$, from the central frequency $\omega_0$. For the derivation of \eqref{eq:propagator}, the spectral frequency is expanded in the following way: $k(\nu)=k(0)+\frac{1}{v_g}\nu+\beta\nu^2$, where $v_g$ stands for the group velocity and $2\beta$ is the group velocity dispersion. As a concrete example we consider SMF28e+ fibre in our analysis, for which $\beta = -1.15 \times 10^{-26} \: \frac{\mathrm{s}^{2}}{\mathrm{m}} $.  We assume that the time reference frame is moving with the wavepackets. Therefore, its center is shifted by $L/v_g$ with respect to a stationary frame.

The spectral wave function of a fiber-coupled SPDC photon pair can  be approximated analytically using a bivariate normal distribution \cite{URen2003, Kolenderski2009, Lutz2014}:
\begin{multline}
\phi(\nu_1,\nu_2)=\frac{1}{\sqrt{\pi } \sqrt{\sigma_1\sigma_2 \sqrt{1-\rho ^2}}}\times\\\times\exp \left(-\frac{1}{2 \left(1-\rho ^2\right)}\left(\frac{\nu_1^2}{\sigma_1^2}+\frac{\nu_2^2}{\sigma_2^2}-\frac{2 \nu_1 \nu_2 \rho }{\sigma_1\sigma_2}\right)\right),
\label{eq:wf:spect}
\end{multline}
where $\sigma_1$ and $\sigma_2$ are the spectral widths of the photons, and $\rho$ is the spectral correlation coefficient. The parameters of the biphoton wave function $\sigma_1$ and $\sigma_2$ and $\rho$ depend on the characteristics of the SPDC source such as the crystal thickness, cut angle, and its dispersion properties \cite{URen2003, Uren2005, Kolenderski2009}. It was recently shown how the spectral correlation $\rho$ can be controlled by adjusting the SPDC setup parameters \cite{Kuzucu2008,Lutz2014,Gajewski2016}. It is noteworthy to say that, in particular, the positive values of spectral correlation can be achieved only in type II phase matching configuration. In the examples presented further on we take $\omega_0=1550$ nm and $\sigma_1=\sigma_2= 1.57$ THz ($2$ nm) if not stated differently. In order to derive the state of the photon pair after the propagation in SMF over the distance $L$, we Fourier-transform the wave function \eqref{eq:wf:spect} into the time domain and apply the propagator introduced in \eqref{eq:propagation}, which results in:
\begin{multline}
\psi_L(t_1,t_2)= \frac{i\sqrt{\sigma_1\sigma_2}\sqrt[4]{1-\rho^2}}{\sqrt{-\pi\left(f(-\sigma_1^2\sigma_2^2)+2 i \beta  L \left(\sigma_1^2+\sigma_2^2\right)\right)}}\times\\\times e^{-\frac{2 i \sigma_1^2\sigma_2^2 \beta L \left(1-\rho ^2\right) \left(t_1^2+t_2^2\right)+\sigma_1^2 t_1^2+\sigma_2^2 t_2^2+2 \sigma_1 \sigma_2 \rho  t_1 t_2}{2f(-\sigma_1^2\sigma_2^2)+4 i \beta  L \left(\sigma_1^2+\sigma_2^2\right)}},
\label{eq:psi:L}
\end{multline}
where we introduced:
\begin{equation}
f(x)=1+4x \beta^2 L^2(1-\rho^2).
\label{eq:def:f}
\end{equation}
\begin{figure}[b]
	\centering
	\includegraphics[width=\columnwidth]{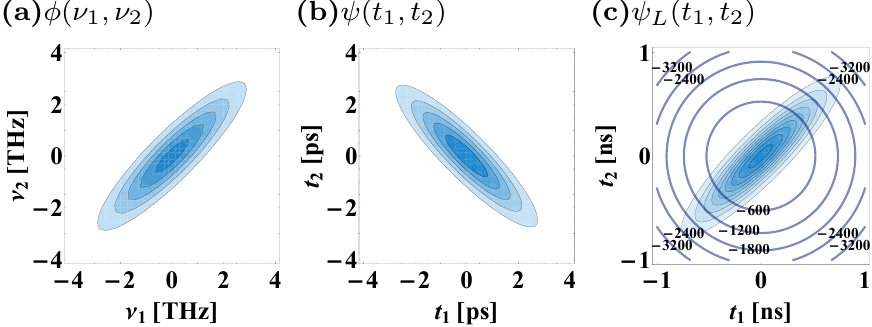}
	%	\begin{tabular}{c c c}
	%		\subfigure[$\psi(\nu_1,\nu_2)$ ]{\includegraphics[width=0.31\columnwidth]{bwni}} &
	%		\subfigure[$\psi(t_1,t_2)$]{\includegraphics[width=0.31\columnwidth]{bwt}} &
	%		\subfigure[$\psi_L(t_1,t_2)$]{\includegraphics[width=0.33\columnwidth]{bwp}} 
	%	\end{tabular}
	\caption{The spectral (a) and temporal (b) wave functions $\phi(\nu_1,\nu_2)$ and  $\psi(t_1,t_2)$ for $L=0$. The functions are real-valued. (c) The phase and the amplitude of the photon pair state, $\psi_L(t_1,t_2)$, after a long propagation distance, $L=10$ km for the initial spectral correlation $\rho=0.9$.}
	\label{fig:psi:prob}
\end{figure}
An example of initial spectral, $\phi(\nu_1,\nu_2)$, and temporal, $\psi(t_1,t_2)$ wave functions, and the state after the propagation are depicted in \figref{fig:psi:prob}. The spectral wave function, $\phi(\nu_1,\nu_2)$, features positive correlation, $\rho=0.9$, which means a positive detuning from the central frequency of photon number 1 is correlated with a positive detuning of photon number 2, see panel (a).  In the time domain it is the opposite as can be seen in panel (b). When the photon pair propagates in SMFs, the type of temporal correlation switches from negative to positive value with increasing distance, see panel (c).

\begin{figure}[b]
	\centering
	\includegraphics[width=\columnwidth]{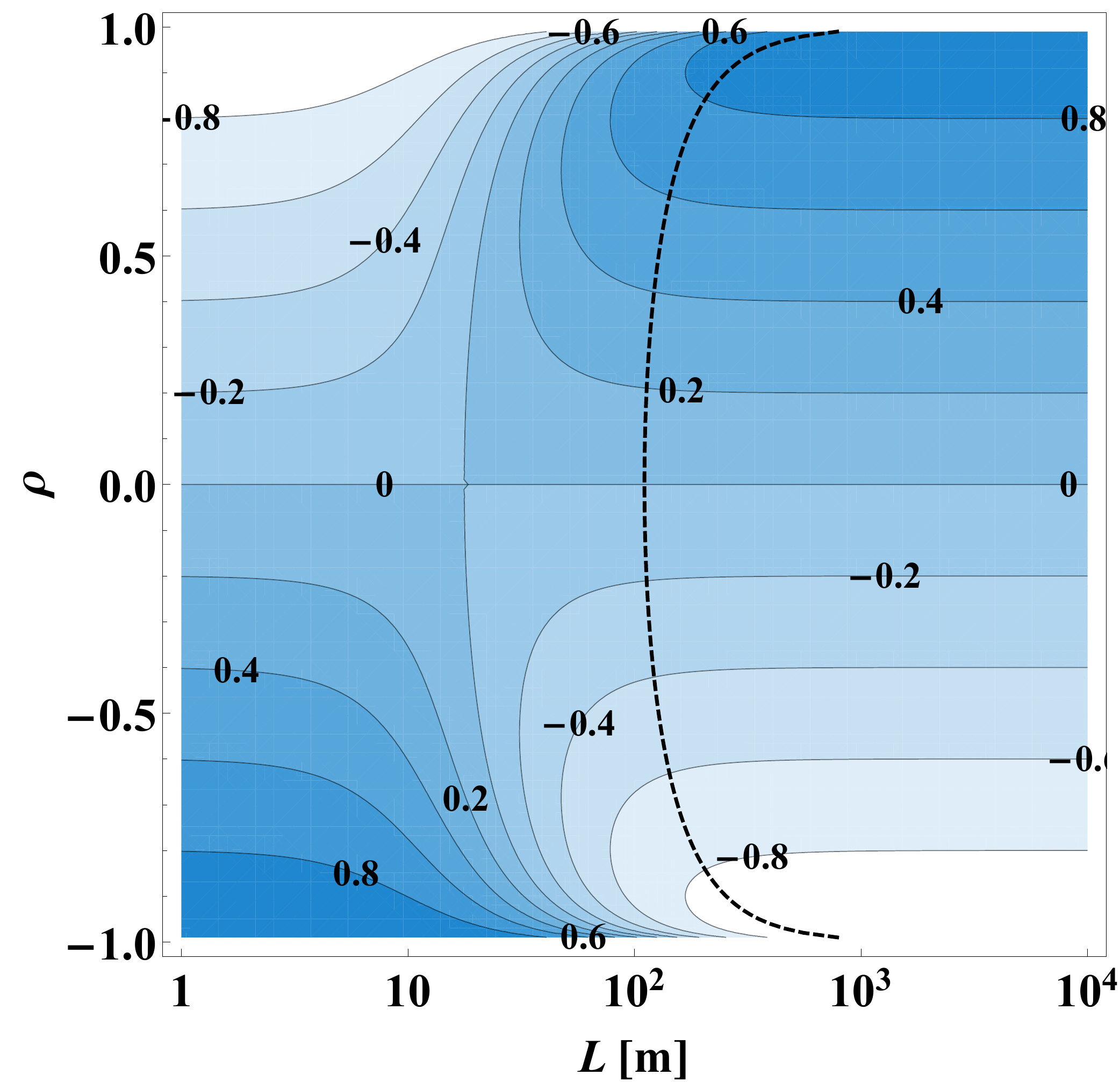} 
	\caption{The Pearson's correlation coefficient, $r_{t_1t_2}(L,\rho)$,  as a function of initial correlation $\rho$ and SMF length $L$. The dashed contour indicates a distance, $L_{0.95}$, at which the correlation coefficient $r_{t_1t_2}(L,\rho) = 0.95 \rho$.  }
	\label{fig:rt1t2}
\end{figure}

\section{Spectral and temporal entanglement}
In order to fully quantify the temporal correlation we resort to the Pearson's coefficient, which has been previously applied in the analysis of SPDC photons in Refs.~\cite{Lutz2014,Gajewski2016}. Note that it is related to the Schmidt number \cite{URen2003, Exter2006}, but carries additional information about the sign of correlation. For the wave function, given in \eqref{eq:psi:L}, it reads:
%is the following function of the propagation distance $L$, and the initial spectral correlation $\rho$:
\begin{equation}
r_{t_1t_2}(L, \rho)=\frac{-\rho f(-\sigma_1^2\sigma_2^2)}{\sqrt{f(\sigma_1^4)f(\sigma_2^4)}}.
\label{eq:rt1t2}
\end{equation}
One can see from \eqref{eq:psi:L} and \eqref{eq:rt1t2} that the correlation in amplitude disappears when $f(-\sigma_1^2\sigma_2^2)=0$ or $\rho=0$, but the correlation in phase disappears only when $\rho=0$. Therefore, if the initial wave function was entangled, it remains such during the propagation process.
It can be also observed that although for the case of $L=0$ the Pearson's coefficient equals $r_{t_1t_2}=-\rho$, for $L\rightarrow \infty$ it transforms into $r_{t_1t_2}=\rho$. This means that in the limit of long fiber lengths the correlation in the time domain is the same as the spectral correlation, which is illustrated in \figref{fig:psi:prob}.

The Pearson's coefficient, $r_{t_1t_2}(L, \rho)$, is depicted in \figref{fig:rt1t2}. From the experimentalist's point of view it is important to know the propagation distance at which the magnitude of the temporal correlation is getting close to the asymptotic limit. Here we consider a distance, $L_{0.95}$, for which the Pearson's coefficient takes the value of $0.95\rho$. It is marked with the dashed contour in \figref{fig:rt1t2}. It can be shown that $L_{0.95}$ gets longer with the decreasing photon bandwidth. As we will see later, an important quantity in the context of the security of long-distance QKD schemes is the distance $L_0$ at which $r_{t_1t_2}=0$. It corresponds to the point where the temporal correlation changes its sign. Utilizing the formulas \eqref{eq:def:f} and \eqref{eq:rt1t2} it is easy to find that:
\begin{equation}
L_0=\frac{1}{2\sigma_1\sigma_2\beta\sqrt{1-\rho^2}}.
\label{eq:L0}
\end{equation}

\begin{figure}[b]
	\centering
	\includegraphics[width=\columnwidth]{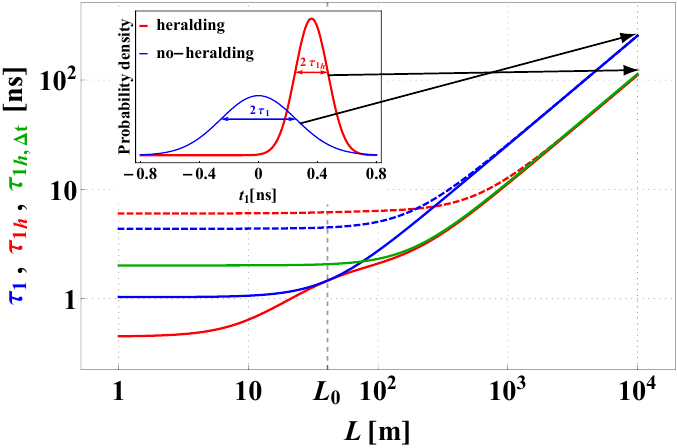}
	%		\begin{overpic}[width=0.8\columnwidth, clip]{deltat1fort2}
	%			\put(15,64.5){
	%			\subfigure{\begin{tabular}{c}
	%			\includegraphics[width=0.32\columnwidth]{probability.pdf}
	%			\end{tabular}}}
	%			\linethickness{0.5pt}
	%			\put(45.75,54.5){\color{black}\vector(3,0.05){49}}
	%			\linethickness{0.5pt}
	%			\put(44.8,47.4){\color{black}\vector(3,0.799){49.5}}
	%		\end{overpic}
	\caption{ The photon detection time widths $\tau_1$ (blue), $\tau_{1h}$ (red) and $ \tau_{1h,\Delta t}$ (green) as a function of the propagation distance $L$ for $\rho=0.9$ and the detection timing jitter $\tau_j=0$ (solid lines) and  $\tau_j=20\,\mathrm{ps}$(FWHM) (dashed lines).  The inset shows the corresponding detection probability functions of finding a  photon at time $t_1$.  }
	\label{fig:exp:t1}
\end{figure}

\section{Detection time characteristics}

The temporal correlation has strong influence on the expected detection time. If information on the detection time of photon number 2 is not available, which we call the no-heralding scenario, the probability density for the detection time of photon number 1 can be computed as a marginal distribution: $\int \mathrm{d} t_2\,\left|\psi_L(t_1,t_2)\right|^2$. In this case the temporal width of photon number 1 can be calculated as the standard deviation of this probability density. It reads:
\begin{equation}
%\Delta t_1=
\tau_1(\sigma_1)=\sqrt{2 \sigma_1^2 \beta ^2 L^2+\frac{1}{2 \sigma_1^2(1-\rho ^2)}}.
\label{eq:tau}
\end{equation}
On the other hand, if the detection time of photon number 2 is known to be $T_2$, the probability density for the detection time of photon number 1 can be calculated as:
\begin{equation}
\frac{\left|\psi_L(t_1,T_2)\right|^2}{\int\mathrm{d}t_1\,\left|\psi_L(t_1,T_2)\right|^2}.
\label{eq:p:con}
\end{equation}
Analogically, in this case the temporal width of photon number 1 is given by:
\begin{equation}
% \Delta t_{1h}=\\
\tau_{1h}(\sigma_1,\sigma_2)=\sqrt{\frac{\left(f(-\sigma_1^2\sigma_2^2)\right)^2+4\beta^2L^2(\sigma_1^2+\sigma_2^2)^2}{2\sigma_1^2f(\sigma_2^4)}}.
\label{eq:tauh}
\end{equation}
We call this the heralding scenario.

\figref{fig:exp:t1} shows the temporal widths for the heralding, $\tau_{1h}$, and no-heralding, $\tau_1$, cases as a function of the propagation distance $L$. It can be seen that heralding results in wave packet narrowing. For long propagation distances, $L$, the ratio $\tau_{1h}(\sigma_1,\sigma_2)/ \tau_{1}(\sigma_1) \approx \sqrt{1-\rho^2}$
%\begin{equation}
% 	\tau_{1h}(\sigma_1,\sigma_2)/ \tau_{1}(\sigma_1) \approx \sqrt{1-\rho^2}
%\end{equation}
depends only on $\rho$. Therefore, the stronger the initial spectral correlation, the narrower the heralded wave packet compared to the not heralded one.
These two temporal widths are equal at the propagation distance $L_0$ given in \eqref{eq:L0}, see \figref{fig:exp:t1}. 

So far we assumed that the detectors' quantum efficiency is $100 \%$ and their timing resolution is perfect, which means that there is no timing jitter. Now we also take into account the possible non-zero detection jitter, $\tau_j$. In this case the overall temporal widths of photon number 1 in the no-heralding and heralding cases can be computed as $\sqrt{\tau_{1h}(\sigma_1,\sigma_2)^2 + 2 \tau_j^2}$ and $\sqrt{\tau_1(\sigma_1)^2 + \tau_j^2}$, respectively. The resulting temporal widths calculated for the cases where $\tau_j=20\,\mathrm{ns}$ (FWHM) are depicted with dashed lines in \figref{fig:exp:t1}. Note that for the (no-)heralding scenario there is the jitter of (one) two detectors taken into account. 

In the situation when photon number 2 was registered at time $T_2$, the center of the probability function for the detection time of photon number 1 moves from $\langle t_1\rangle=0$ to 
\begin{equation}
\langle t_1\rangle_{T_2} ={-T_2  \rho \sigma_2  f(-\sigma_1^2\sigma_2^2)}/{(\sigma_1 f(\sigma_2^4))}.
\end{equation}
This means that in order to get the advantage from the possibility to reduce the duration time of her detection windows, Alice has to shift the center of this window accordingly to the above formula for each detection event at Bob's setup individually. Otherwise the advantage is lost. The proportionality factor depends on the initial spectral correlation $\rho$.

In many experimental applications the exact detection time of the heralding photon may be unknown due to \emph{e.g.} the lack of global time reference.  Let us assume that only the difference between the detection times of the two photons, $\Delta t = t_2 -t_1$, can be measured (i.e. Alice and Bob have only the mutual time reference, while the global time reference is not available to them). Starting from the formula \ref{eq:p:con} expressed in terms of $t_1$ and $\Delta t$ it can be shown that the temporal width of the heralded photon, calculated as the standard deviation of the probability distribution for the time difference $\Delta t $, reads:
\begin{equation}
%\small
\tau_{1h,\Delta t}(\sigma_1,\sigma_2)=\sqrt{\frac{(\sigma_1^2+\sigma_2^2)f\left(\sigma_1^2\sigma_2^2\right)+2\sigma_1\sigma_2\rho f\left(-\sigma_1^2\sigma_2^2\right)}{2 \sigma_1^2\sigma_2^2(1-\rho^2)}}.
\label{eq:tau2h}
\end{equation}
The above temporal width is plotted with the green solid line in \figref{fig:exp:t1}. From this figure it is clear that even if the global time reference is not distributed to Alice and Bob, it can be possible to narrow the temporal width of the heralded photon as compared to the no-heralding case, if only the propagation distance is long enough. This effect is essential for the QKD security analysis performed in the further section.

In all three of the cases analyzed in our work the temporal widths of the photon number 1 depend on the value of $\rho$. Therefore, by manipulating the type of spectral correlation between the photons produced in the SPDC source it can be possible to narrow these temporal widths. This possibility can be clearly seen in \figref{fig:det-times}, presented in \appref{app:C}. This effect has a lot of potential applications. In particular, it can allow for the decreasing of the duration time of the detection windows for the signal and idler photons. In this way one can improve the efficiency of the temporal filtering, which is a popular method used for the reduction of the detection noise in various quantum communication protocols.

\section{Application}
\begin{figure}[t]
	\centering
	\includegraphics[width=\columnwidth]{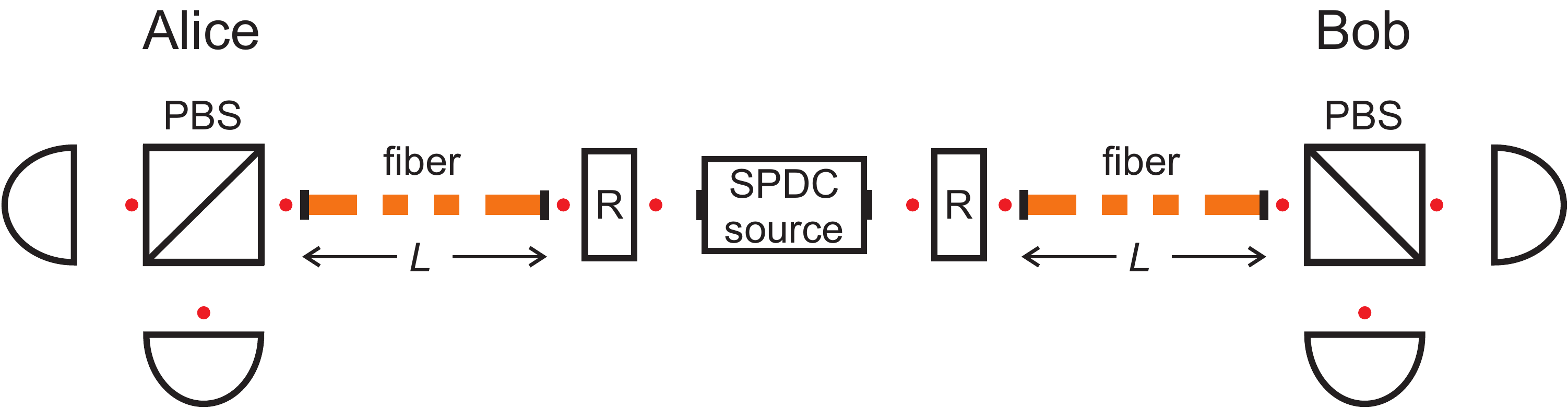}
	\caption{A source-in-the-middle scheme for the implementation of DV QKD protocols. R denotes polarization rotators.}
	\label{fig:QKDscheme}
\end{figure}

As an example, the application that can benefit from narrowing the temporal width of the SPDC photons we consider a standard DV QKD scenario depicted in \figref{fig:QKDscheme}. The detailed description of the scheme and its security analysis are included in \appref{app:A}. In this section of our article we will demonstrate how the maximal security of this scheme can be increased by manipulating the type of spectral correlation between the photons produced by the SPDC source and subsequently sent to Alice and Bob. We assume here that Alice and Bob use four identical free-running detectors with the dark count rate $d=10^3\,\mathrm{counts/s}$, quantum efficiency of $100\%$, timing jitter $\tau_j$ and without the photon-number-resolution ability. We also assume that the source  produces exactly one pair of photons at a time. The loss related to  propagation in  SMF28e+ is taken into account. 
% described by the attenuation coefficient $\alpha=0.2\,\mathrm{dB/km}$. Thus, their transmittance can be calculated as $T=10^{-\alpha L/10}$.

We consider two situations for the application of temporal filtering: i) global time reference is available for Alice and Bob, which means that they always know the exact moments $t_1$ and $t_2$ of their clicks in relation to the time of the generation of a given photon pair, and ii) global time reference is not available, but Alice and Bob share mutual time reference, so they do not know $t_1$ and $t_2$ but can find the value of relative time difference between their detections, $\Delta t=t_2-t_1$. 

The temporal filtering for the first case, i), consists of two parts. Firstly, the global time reference allows Alice and Bob to reduce their dark count rates even without communicating with each other. It is because for each photon pair they can separately find time windows in which they can expect their detection to take place and disregard all of the other clicks. We assume that Alice fixes the duration of her time windows on the level of $\tau_1^I(\sigma_1)=6 \tau_1(\sigma_1)$, which gives her the probability of $99.73\%$ for a successful detection. Analogously, the duration of Bob's time windows equals $\tau_1^I(\sigma_2)$. In principle, Alice and Bob could establish shorter detection windows in order to reduce the number of registered dark counts even more at the expense of losing considerable part of the SPDC photons. In the idealized situation considered here, in which all of the errors appearing in the protocol originate from the dark counts and the SPDC source never emits more than one pair of photons at a given time, shortening the detection window would always lead to the increase of the maximal security distance for QKD. However, key generation rate would be reduced at the same time, which is not desirable from the practical point of view. Moreover, all of the results presented in this paper are qualitatively the same if we consider shorter detection windows for Alice and Bob. 

The second part of the temporal filtering can be performed during the basis reconciliation stage when Alice and Bob communicate with each other and disregard all of their measurement results that cannot be paired up properly. To do this, one of them, say Bob, sends to the other one all his detection times. Next, for each one of these time moments Alice assigns the corresponding time window, which equals to $\tau_{1h}^I(\sigma_1,\sigma_2)=6\tau_{1h}(\sigma_1,\sigma_2)$, where she searches for the click registered by her own detector. Only if Alice's search was successful, a given pair of clicks is accepted. The roles of Alice and Bob can be interchanged. 
%If Bob is the person who applies temporal filtering to his measurement results, knowing Alice's detection times, the width of his time windows should be equal to $\tau_{2h}^I(\sigma_1,\sigma_2)=\tau_{1h}^I(\sigma_2,\sigma_1)$.

On the other hand, if the global time reference is not available, the legitimate participants cannot apply temporal filtering for their results separately. In order to calculate the dark count rates, we define  $\tau_{1}^{II}$ as the duration of a time slot corresponding to a single photon pair.  Since the whole set of all such time slots have to cover the entire duration of the key generation process, $\tau_{1}^{II}$ can be related to the repetition rate of the source as $\tau_{1}^{II}={1}/{r}$. In practice, $r$ can be limited by various experimental factors \cite{Scarani09}. If the photons are transmitted over a long distance, the dispersion effects also limit the maximal repetition rate,  because of  an overlap of the time slots of two consecutive signals. It can increase the ratio of errors and deteriorate the key. We do not optimize over the repetition rate, but we assume it is equal to $r=10\mathrm{MHz}$, which is a reasonable value for typical realistic implementations.

Even if the global time reference is not available to them, but Alice and Bob share mutual time reference, they can apply temporal filtering during the stage of basis reconciliation. After receiving information from Bob on the time moments at which he registered all of his clicks, Alice should set the time windows for her own measurement results based on the function of the detection probability derived for $\Delta t$. The optimal width for such window is equal to $\tau_{1h}^{II}(\sigma_1,\sigma_2)=6\tau_{1h,\Delta t}(\sigma_1,\sigma_2)$. Once again the roles of Alice and Bob can be interchanged.
% When Bob is the one searching for clicks which could be paired up with Alice's clicks, the width of time windows considered by him should be equal to $	\tau_{2h}^{II}(\sigma_1,\sigma_2)=\tau_{1h}^{II}(\sigma_2,\sigma_1)$.

\begin{figure}[b]
	\centering
	\includegraphics[width=0.9\columnwidth]{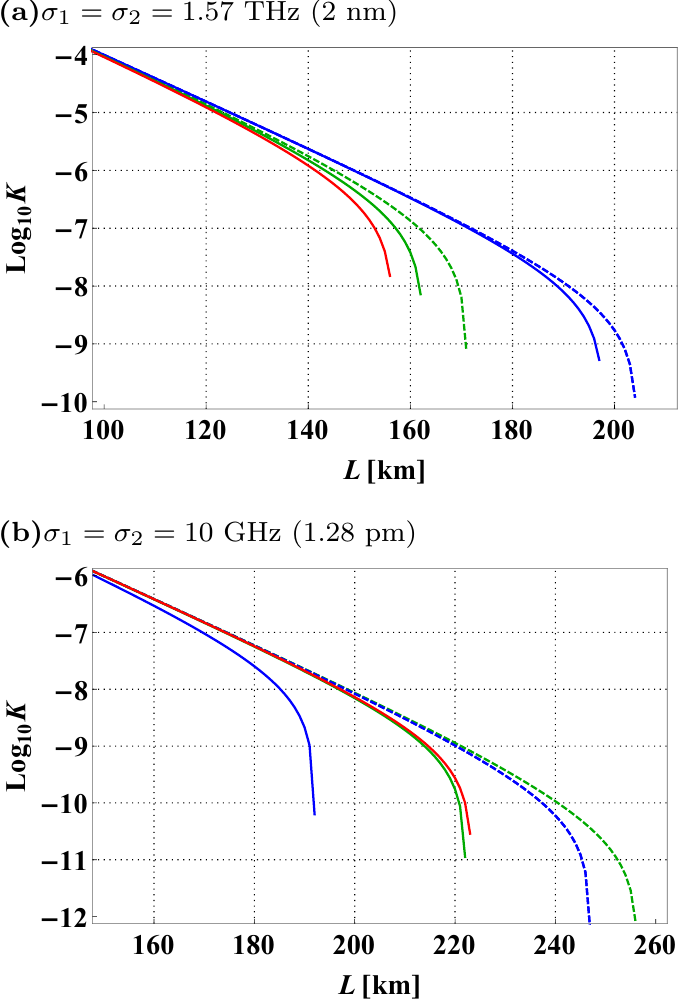}
	%	\begin{tabular}{c }
	%		\subfigure[$\sigma_1=\sigma_2=1.57$ THz ($2$ nm)]{\includegraphics[width=0.8\columnwidth]{correlations1a}} \\
	%		\subfigure[$\sigma_1=\sigma_2=10$ GHz ($1.28$ pm)]{\includegraphics[width=0.8\columnwidth]{correlations1b}}
	%	\end{tabular}
	%	\begin{tabular}{c c }
	%	\subfigure[ ]{\includegraphics[width=0.45\columnwidth]{correlations1a}} &
	%		\subfigure[]{\includegraphics[width=0.45\columnwidth]{correlations1b}}
	%	\end{tabular}
	\caption{Key generation rate, $K$, as a function of the distance between the source and the legitimate participants of the BB84 protocol for photon bandwidths a) $\sigma_1=\sigma_2=1.57$ THz and b) $\sigma_1=\sigma_2=10$ GHz, and spectral correlation: $\rho=-0.9$ (red lines), $\rho=0$ (green lines) and $\rho=0.9$ (blue lines). All the plots were made for the case of the source repetition rate  $r=10\,\mathrm{MHz}$ and detection timing jitter $\tau_j=0$. The dashed (solid) lines correspond to the case when the global time reference is (is not) available for Alice and Bob. The dashed red lines are not visible on both of the panels since they are identical to the respective dashed blue lines.}
	\label{fig:QKDresults}
\end{figure}

Figure \ref{fig:QKDresults} presents the results of our security analysis of the QKD scheme for the situations i) and ii) for the BB84 protocol \cite{Bennet1984} with the assumption that the detectors used by Alice and Bob are not affected by the detection jitter. By looking at the panel (a) one can conclude that the security of the DV QKD protocols can be  significantly improved by utilizing photon pairs featuring positive spectral correlation when the information about the global time reference is not available to Alice and Bob. In the opposite case, what matters is not the type of spectral correlation, but its strength, as the maximal security distance grows when $|\rho|$ increases.

Note that the above conclusions can be substantially different for smaller bandwidths $\sigma_1$ and $\sigma_2$. It can be seen in \figref{fig:QKDresults} (b), where a plot analogous to the one in panel (a) was made for $\sigma_1=\sigma_2=10$ GHz. First of all, for the case when the global reference time is not available to Alice and Bob, positive spectral correlation appears to be worse than the negative one in the context of QKD security. This can be explained by the fact that the value of $L_0$, for which the initial type of temporal correlation switches to the opposite one, strongly depends on the bandwidth as given in \eqref{eq:L0}.  For $\sigma_1=\sigma_2=10$~GHz this distance is already much longer than the maximal security distance for the BB84 protocol. Therefore, in this case at the maximal security distance negative spectral correlation still corresponds to the positive temporal correlation (as for $L=0$) and vice versa. This is opposite to the case of $\sigma_1=\sigma_2=1.57$ THz, plotted in panel (a).

\begin{figure}[h]
	\includegraphics[width=0.95\columnwidth]{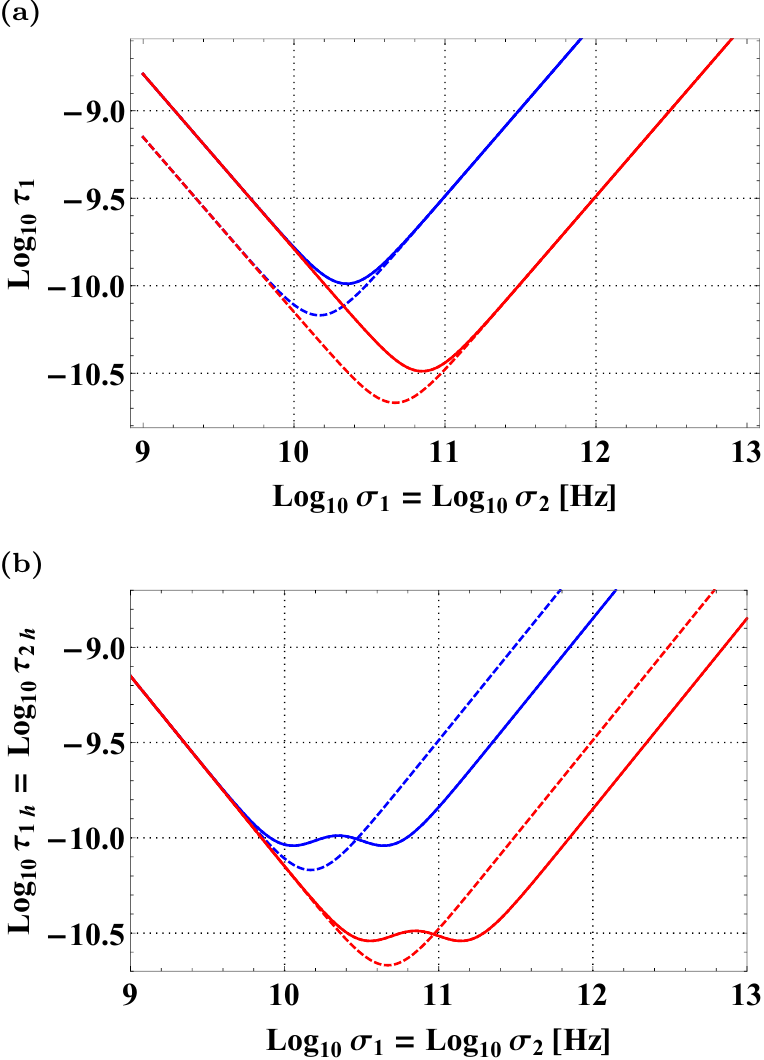}
	\caption{Temporal widths a) $\tau_1$ and b) $\tau_{1h}=\tau_{2h}$, given by the formulas (\ref{eq:tau}) and (\ref{eq:tauh}) respectively, plotted as functions of spectral widths $\sigma_1=\sigma_2$ for the spectral correlation coefficient $\rho=0$ (dashed lines) and $\rho=0.9$ (solid lines) and for two different lengths of the standard SMF fibers connecting the source with Alice and Bob: 20 km (red lines) and 200 km (blue lines). All the plots were made for the detection timing jitter $\tau_j=0$.}
	\label{fig:TemporalWidths}
\end{figure}

There is another surprising conclusion which can be drawn from \figref{fig:QKDresults} (b). In the case when the global time reference is available to Alice and Bob the lack of spectral correlation between the photons may be more advantageous for the participants of a given QKD protocol than the strong correlation. This can be explained by looking at the \figref{fig:TemporalWidths} where the values of the temporal widths $\tau_1$, $\tau_{1h}$ and $\tau_{2h}$ were plotted as functions of the spectral widths $\sigma_1=\sigma_2$ for different values of $L$ and $\rho$. By looking at the left-hand side of the panels (a) and (b) of this figure, one can conclude that for narrow spectral widths of the photons emitted in the SPDC process the values of $\tau_{1h}$ and $\tau_{2h}$ are fairly independent of $\rho$. On the other hand $\tau_1$ is significantly lower for $\rho=0$ than for $|\rho|=0.9$. Therefore, the combined influence of these temporal widths on the key generation rate (described in details in \appref{app:B}) has to be more favorable in the case of weak spectral correlation. However, when the spectral widths of the photons emitted in the SPDC process are relatively broad (see the right-hand side of the panels (a) and (b) in \figref{fig:TemporalWidths}), the value of the temporal width $\tau_1$ becomes independent of $\rho$, while the values of $\tau_{1h}$ and $\tau_{2h}$ become smaller for $|\rho|=0.9$ than for $\rho=0$. Thus, in this case the combined influence of $\tau_1$, $\tau_{1h}$ and $\tau_{2h}$ on the key generation rate is certainly more favorable in the case of strong spectral correlation.

It is also worth noting that in all of the cases considered in \figref{fig:QKDresults} the maximum security distance that can be read out from the panel (b) turns out to be significantly longer than the corresponding maximum security distance in the panel (a). The reason for this is that the narrower are the spectral widths of the photons created in SPDC process, the smaller are the effects of temporal broadening introduced by the chromatic dispersion during the propagation of these photons in the telecommunication fibers. Therefore, the temporal widths of the signal and idler photons at the output of these fibers are generally smaller for the case of narrow spectral widths. 

However, one should also be aware of the fact that in realistic situation strong narrowing of the spectral widths of the photons produced in SPDC process would certainly decrease the efficiency of a given source. This fact, which is not included in our simplified analysis, would have negative influence on the key generation rate. In order to evaluate the ultimate result of the two effects described above on the QKD security it would be necessary to perform more elaborate investigation. It can be predicted that in order to maximize the security distance of a given QKD scheme based on SPDC source in realistic situation, Alice and Bob would probably have to perform some optimization of the spectral bandwidth of the SPDC photons. The investigation of this problem lies outside of the scope of this paper.

%\begin{figure}[h]
%	\centering
%	\includegraphics[width=0.95\columnwidth]{Figure7}
%\caption{Key generation rate, $K$, as a function of the distance between the source and the legitimate participants of the BB84 protocol for photon bandwidths  $\sigma_1=\sigma_2=10$ GHz, spectral correlation $\rho=-0.9$, source repetition rate  $r=10\,\mathrm{MHz}$ and timing jitter for all the detectors equal to $\tau_j=0$ (blue lines) or $\tau_j=500\,\mathrm{ps}$ (orange lines). The dashed (solid) lines correspond to the case when the global time reference is (is not) available for Alice and Bob.}
%\label{fig:QKDjitter}
%\end{figure}

While the results presented in \figref{fig:QKDresults} were obtained with the assumption that $\tau_j=0$, in realistic situation non-zero timing jitter of the detectors may considerably influence the security. We investigated this problem and found out that the level of $\tau_j$ above which the effect of timing jitter on the QKD security stops being negligible strongly depends on the spectral widths of the photons generated in SPDC process. On one hand, for $\sigma_1=\sigma_2=1.57$ THz the negative influence of the timing jitter on the key generation rate can be safely neglected in every scenario as long as $\tau_j<200\,\mathrm{ps}$. This inequality can be fulfilled not only for the high-quality superconducting detectors \cite{Rosfjord2006,Takesue2007,Divochiy2008}, but also for InGaAs/InP detectors \cite{Dixon2008,Tosi2012a}. However, when $\sigma_1=\sigma_2=10$ GHz and the global time reference is available to Alice and Bob the effect of the timing jitter can become visible even for $\tau_j=50\,\mathrm{ps}$. This is relatively low value, which can be reached only in some high-quality superconducting detectors. More detailed results of our investigation of the influence of the detection timing jitter on the QKD security can be found in the \appref{app:B}.

\section{Conclusions}
In conclusion, we analyzed the state of SPDC pair propagated through two identical SMFs introducing chromatic dispersion. This allowed us to analytically investigate the temporal characteristics of these photons. In particular, we analyzed the possibility to reduce the detection noise by applying the procedure of temporal filtering. The results were used to evaluate the security of a DV QKD scheme with the SPDC source located in the middle between Alice and Bob. We analyzed two situations depending on the availability of the global time reference. We showed that when it is not available the maximal security distance can be increased by utilizing the source with positive spectral correlation, if the spectral bandwidth of this source is relatively broad. A similar effect is not possible when the global time reference is  available for Alice and Bob. In this case the security depends only on the strength of the spectral correlation but not on its sign. We also discovered that for narrow-band SPDC photons, the aforementioned advantage of the positive correlation over the negative one can be reversed, while the source emitting uncorrelated photons can be the best one in the context of QKD security.

\section*{Funding Information}
The authors acknowledge support by the National Laboratory FAMO in Torun, Poland, financial support by Foundation for Polish Science under Homing Plus no.~2013-7/9 program supported by European Union under PO IG project and by Polish Ministry of Science and Higher Education under grant 6576/IA/SP/2016 and Iuventus Plus grant no.~IP2014 020873.

%\bibliographystyle{apsrev4-1}
%\bibliography{FamoLab3}


\appendix
\section{A Discrete Variable Quantum Key Distribution scheme}\label{app:A}

From each pair of photons generated by the SPDC source in the DV QKD scheme pictured in Fig.~4 one photon is subsequently sent to Alice and the other one to Bob. Before being coupled in SMFs of lengths $L$, both of them go through similar polarization rotators, which always rotate their polarizations by the same angle, randomly chosen to be $0^\circ$ or $45^\circ$. After receiving the signal, Alice and Bob independently perform polarization measurements on their respective photons by using standard setups, consisting of two single-photon detectors and a polarizing beam-splitter. The beam-splitters can be rotated about the axis parallel to the direction of propagation of the photons sent from the source. By doing this the legitimate participants of a given QKD protocol can change the bases for their measurements. In the ideal situation, if only Alice and Bob chose the same basis and this basis is compatible with the rotation applied to the photons by the polarization rotators, they can be sure that the results of their measurements are perfectly correlated with each other. In this case the key generated by the scheme illustrated in Fig.~4 is totally secure.

The most basic quantity used to describe the security of the QKD protocols in realistic situations is the key generation rate. Here it is defined as the number of secure bits of the key that can be distilled by Alice and Bob per one pair of photons created by the SPDC source. It can be expressed as \cite{Scarani09,Devetak05}:
\begin{equation}
K=p_\mathrm{exp}\max\left[0,I_{AB}-\min\left\{I_{EA},I_{EB}\right\}\right],
\label{Eq:KeyRateMain}
\end{equation}
where in the case of the setup depicted in Fig.~4, $p_\mathrm{exp}$ denotes the probability that after the emission of a single pair of photons by the SPDC source both Alice and Bob get at least one click in one of their detectors and accept a given event for the process of key generation. Next, $I_{AB}$ is the mutual information about the generated raw key shared by Alice and Bob and $I_{EA}$ ($I_{EB}$) describes the amount of information on Alice's (Bob's) version of the raw key which a spy can gain upon an eavesdropping attack. Since the analyzed setup is symmetric, we can assume that $I_{EA}=I_{EB}$. For definiteness from now on we consider Alice's version of the key as the base version, information on which Eve and Bob are trying to gain. For the BB84 protocol \cite{Bennet1984}, which we  focus on in our calculations, the upper bound on $I_{EA}$, which Eve can get by performing the most general collective attacks, equals \cite{Kraus2005}:
\begin{equation}
I_{EA}^\mathrm{BB84}=H(Q),
\end{equation}
where 
\begin{equation}
H(Q)=-Q\log_{2}Q-(1-Q)\log_2(1-Q)
\end{equation}
is the Shannon entropy and $Q$, called quantum bit error rate (QBER), represents the ratio of errors in Bob's version of the raw key. On the other hand, the mutual information between Alice and Bob can be expressed as \cite{Scarani09}:
\begin{equation}
I_{AB}=1-H(Q).
\end{equation}
In the end, from the formula (\ref{Eq:KeyRateMain}) for the BB84 protocol we get
\begin{equation}
K=p_\mathrm{exp}\max\left[0,1-2H(Q)\right].
\label{Eq:KeyRateTrans}
\end{equation}

From \eqref{Eq:KeyRateTrans} we can see that the key generation rate depends on two quantities: the probability of accepting a given event by Alice and Bob for the process of key generation and the ratio of errors in Bob's version of the key. For any specific implementation of the BB84 protocol both of these two quantities can be expressed in terms of the parameters of a given setup. In the case of the scheme presented in Fig.~4,  $p_\mathrm{exp}$ can be approximated by:
\begin{equation}
p_\mathrm{exp}\approx T^2+T(1-T)\left[P_{1h}+P_{2h}\right]+(1-T)^2P_1P_{2h},
\label{eq:pexp}
\end{equation}
where $P_{1h}\approx2d\tau_{1h}$ is the probability of registering a dark count by Alice's measurement system during the time window $\tau_{1h}$ and the other probabilities $P_x$ can be defined in the analogous way. Note that $\tau_{2h}$ needed for the calculation of $P_{2h}$ can be related to $\tau_{1h}$ as follows: $\tau_{2h}(\sigma_1,\sigma_2)=\tau_{1h}(\sigma_2,\sigma_1)$. The losses of photons related to  propagation in SMFs (for SMF28e+) described by the attenuation coefficient $\alpha=0.2\,\mathrm{dB/km}$. Thus, their transmittance can be calculated as $T=10^{-\alpha L/10}$. The first term in \eqref{eq:pexp} represents the probability that both photons created by the SPDC source in a single attempt and subsequently sent to Alice and Bob are detected by their respective measurement systems. The second term gives us the information about the probability that only one of these photons is detected, but during the corresponding time window a dark count is registered in the other measurement system. Finally, the last term in \eqref{eq:pexp} represents the probability that both signal photons are lost during their transmission in the fibers.  However, Alice and Bob are able to find a pair of dark counts, which could be mistakenly taken for the clicks caused by these photons and therefore accepted for the process of key generation. The approximation \eqref{eq:pexp} is valid if only all the probabilities $P_x$ appearing in this formula are much smaller than unity. This is always  the case for the realistic values of dark count rates.

Since the dark counts appear in the detectors of Alice and Bob totally randomly, there is always a $50\%$ chance for an error in Bob's version of the key when at least one of the signal photons emitted by the SPDC source is lost, but a given event is still accepted by Alice and Bob for the process of key generation. On the other hand, if both signal photons reach the measurement systems of the legitimate participants, they can be sure that the results obtained by them are perfectly correlated. If so, the value of QBER can be calculated using the formula:
\begin{equation}
Q=\frac{p_\mathrm{exp}-T^2}{2p_\mathrm{exp}}.
\label{eq:qber}
\end{equation}

In order to estimate the security of the BB84 protocol for the setup configuration illustrated in Fig. 4 in the case when the information about the time reference of the SPDC source is (is not) distributed to Alice and Bob one has  to find the values of $p_\mathrm{exp}$ and $Q$ given  by \eqref{eq:pexp} and \eqref{eq:qber} respectively and subsequently use the formula \eqref{Eq:KeyRateTrans} to calculate the key generation rate. 

\section{The influence of the timing jitter of the detectors on key generation rate}\label{app:B}

\begin{figure}[t]
	\centering
	\begin{tabular}{c}
		\subfigure[]{\includegraphics[width=0.49\columnwidth]{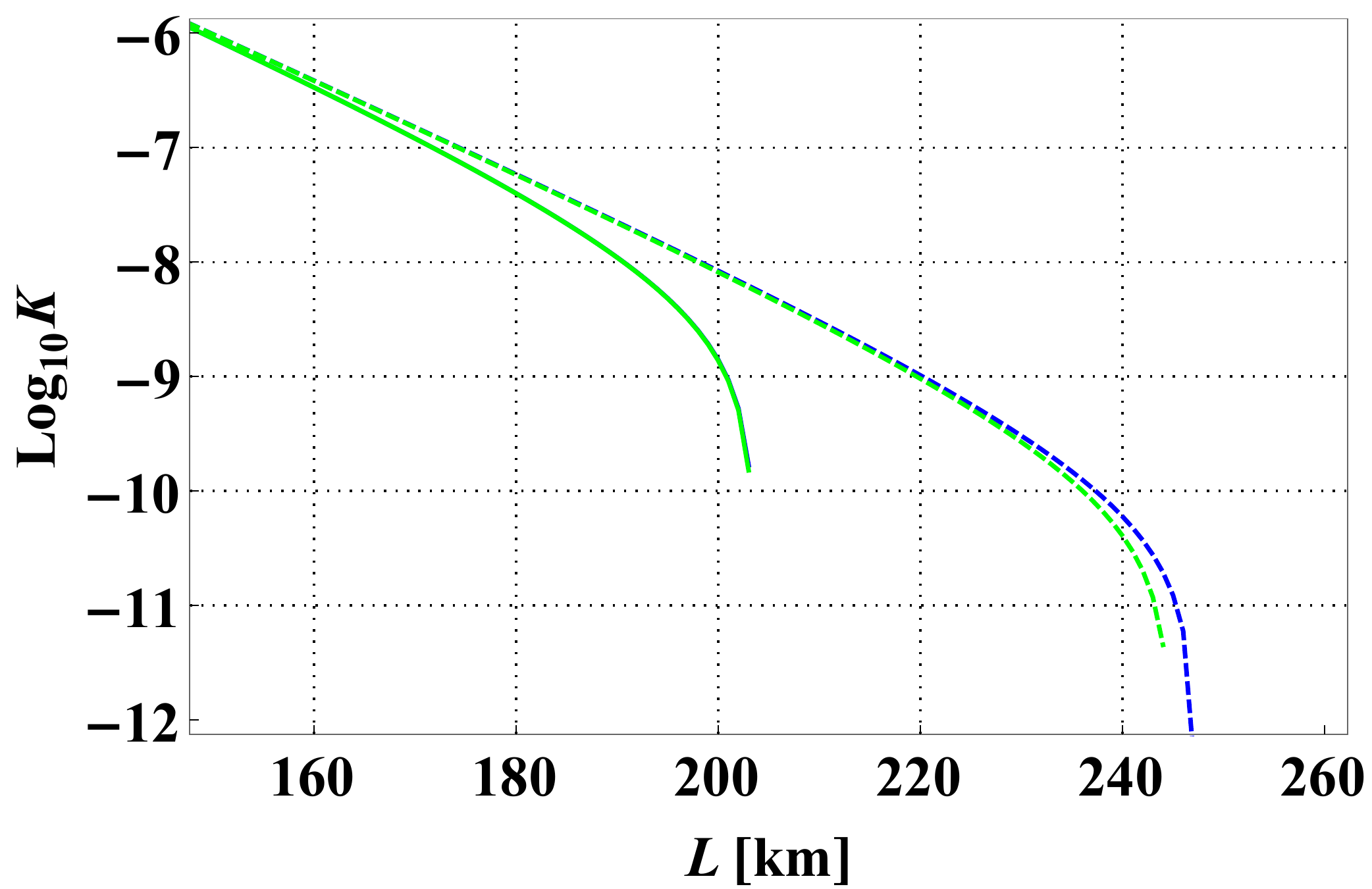}}
		\subfigure[]{\includegraphics[width=0.49\columnwidth]{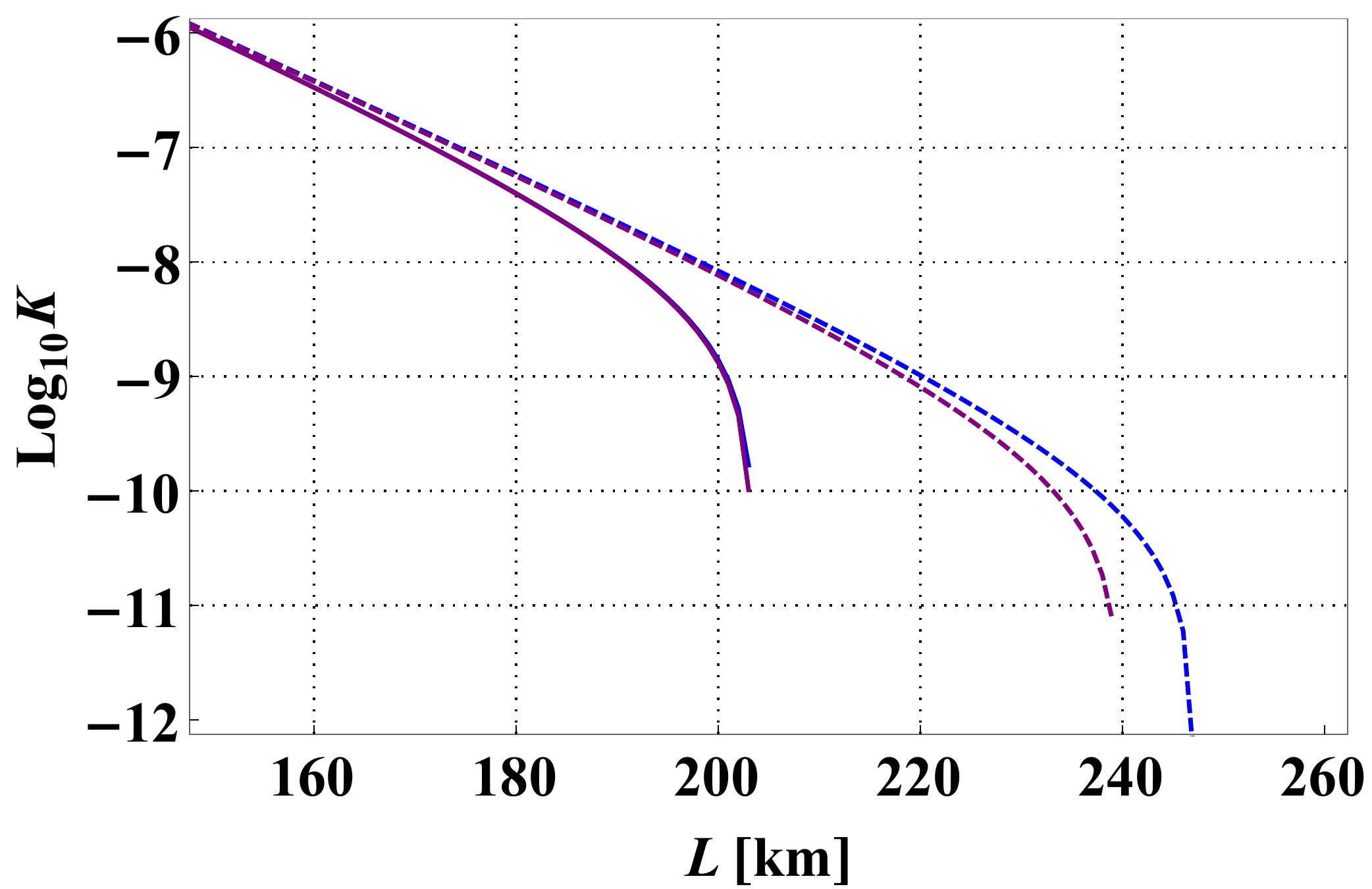}}\\
		\subfigure[]{\includegraphics[width=0.49\columnwidth]{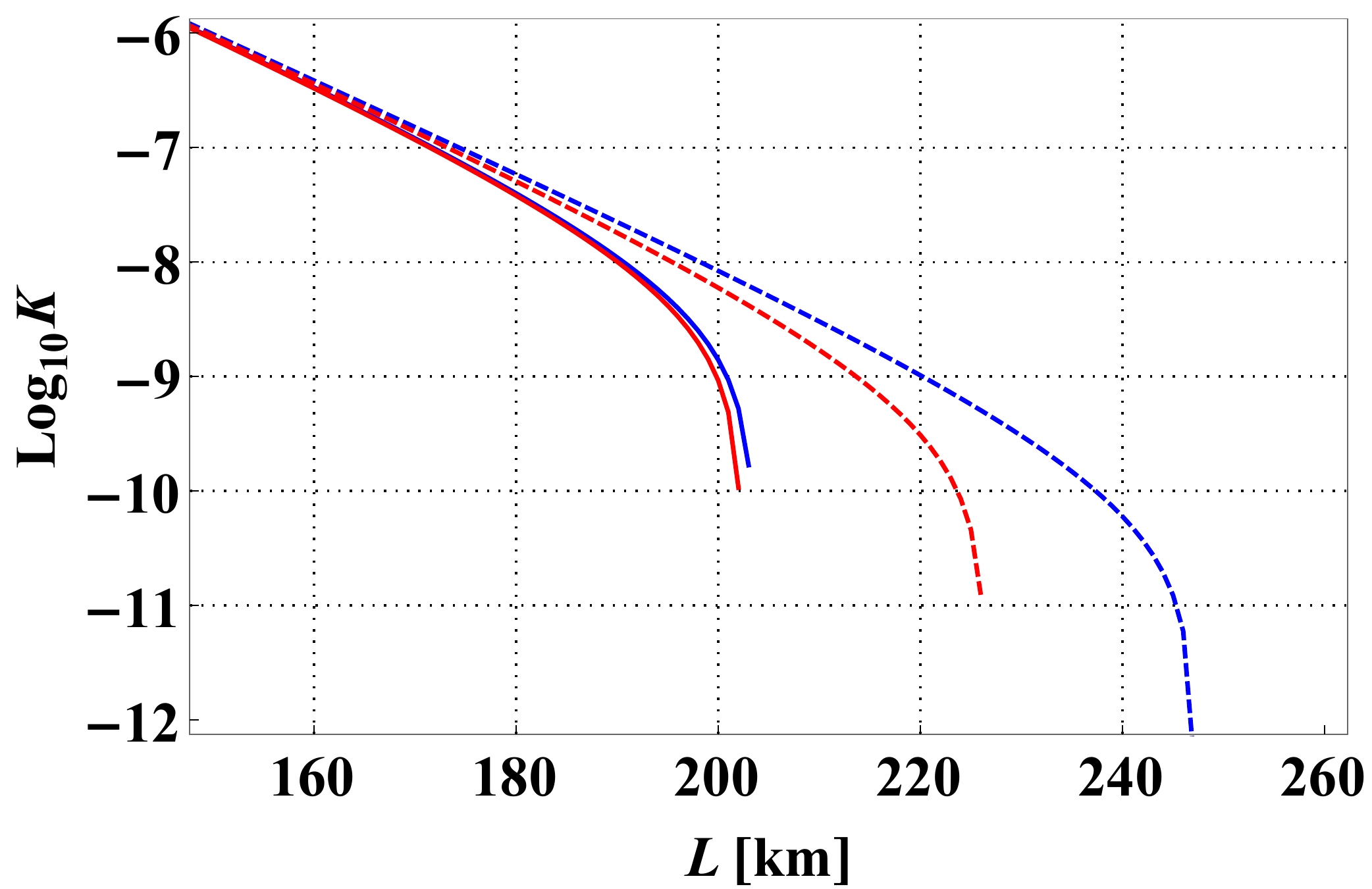}}
		\subfigure[]{\includegraphics[width=0.49\columnwidth]{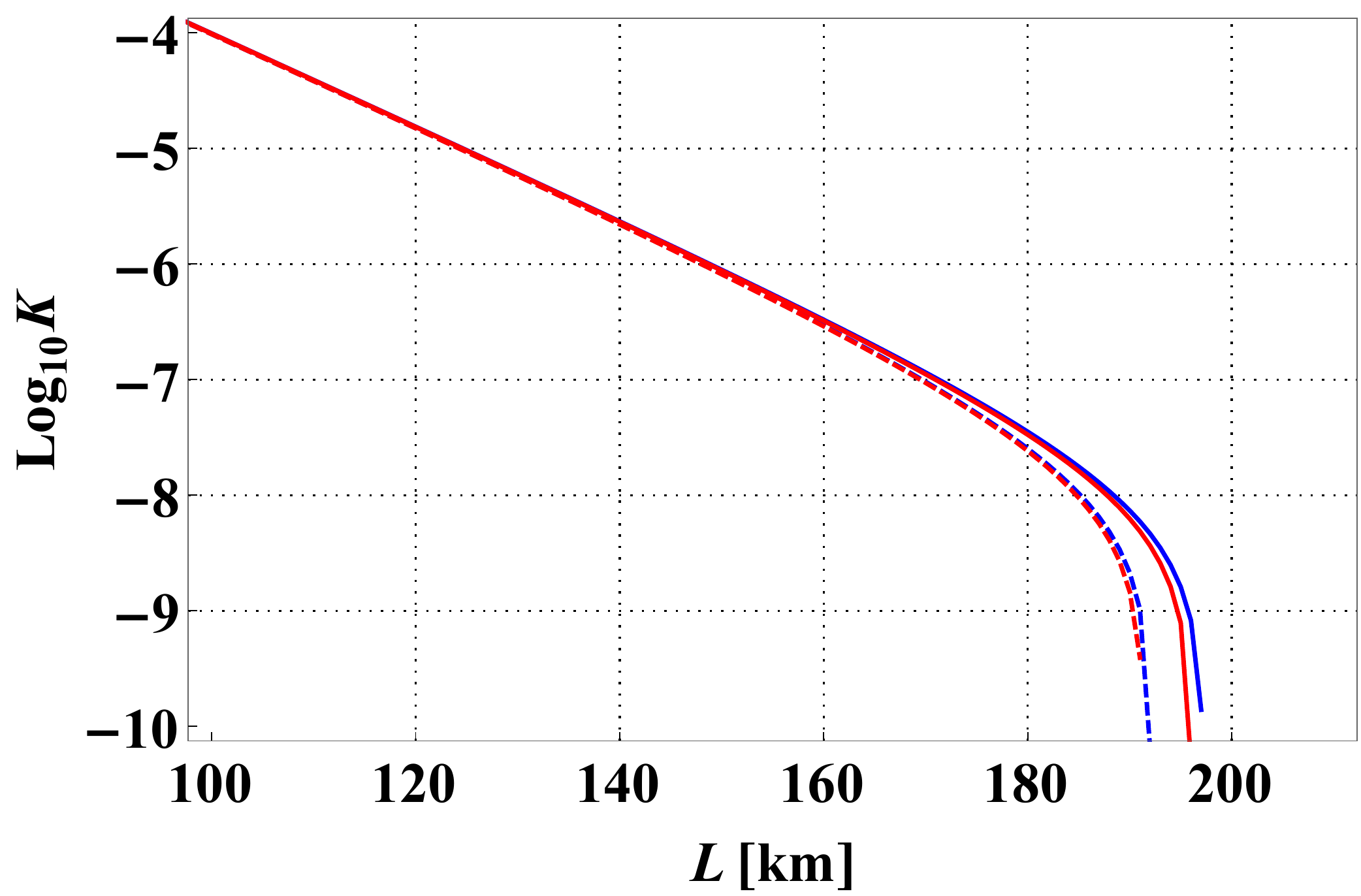}}\\
		\subfigure[]{\includegraphics[width=0.49\columnwidth]{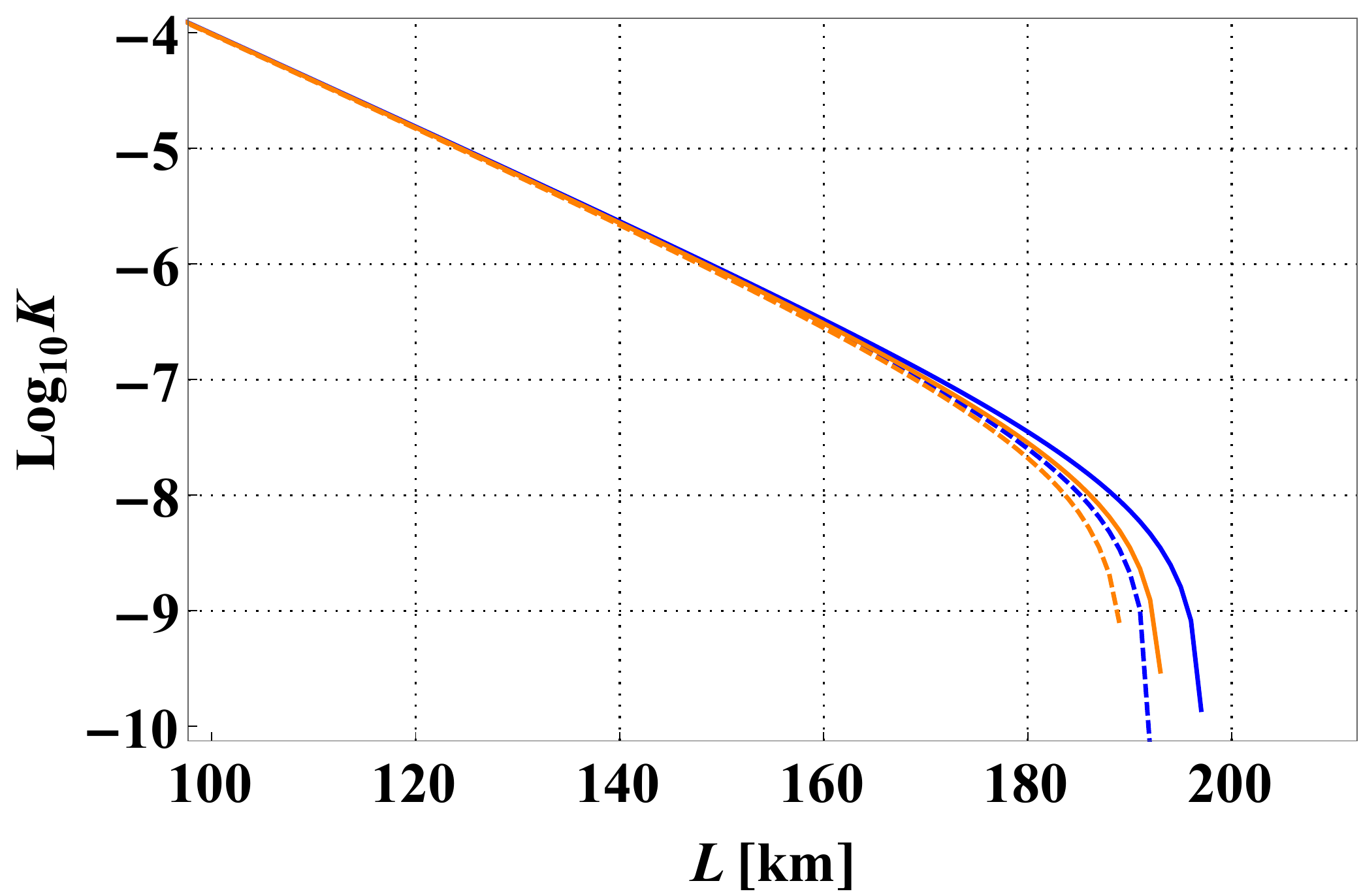}}
		\subfigure[]{\includegraphics[width=0.49\columnwidth]{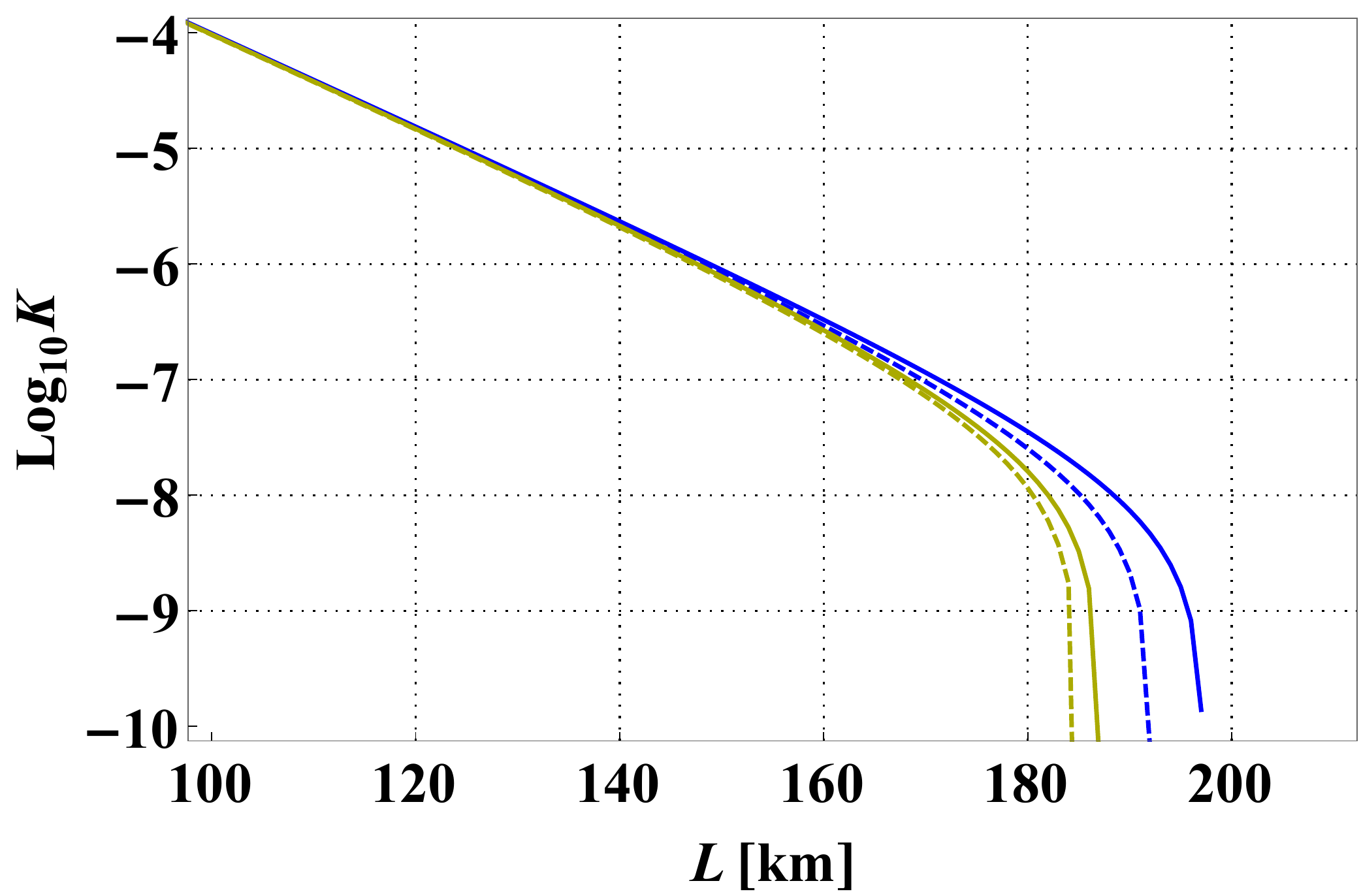}}
	\end{tabular}
	
	\caption{Key generation rate, K, as a function of the distance between the source and the legitimate participants of the BB84 protocol drawn for $\rho=0.99$ in the cases when the global time reference is available (panels (a)-(c)) or is not available (panels (d)-(f)) and the detection jitters are $\tau_j=0$ (blue lines), $\tau_j=0.1\,\mathrm{ns}$ (green lines), $\tau_j=0.2\,\mathrm{ns}$ (purple lines), $\tau_j=0.5\,\mathrm{ns}$ (red lines), $\tau_j=1.0\,\mathrm{ns}$ (orange lines) and $\tau_j=2.0\,\mathrm{ns}$ (yellow lines). The solid (dashed) lines denote the results obtained for the case of $\sigma_1=\sigma_2=1.57\,\mathrm{THz}$ ($\sigma_1=\sigma_2=10\,\mathrm{GHz}$). The results presented on panels (d)-(f) were obtained for the repetition rate of the SPDC source r = 10MHz.}
	\label{fig:Qjitter2}
\end{figure}

All of the results of the QKD security analysis presented in this article were obtained with the assumption that the detectors used by Alice and Bob are not affected by the timing jitter. In order to find conditions for the validity of this assumption, we calculated the function of key generation rate for a few different values of $\tau_j$ in several QKD scenarios and compared it with the analogous function calculated for the ideal case of $\tau_j=0$. A series of plots illustrating the comparison was pictured in \figref{fig:Qjitter2}.

One can conclude that the key generation rate is the most sensitive to the change of value of $\tau_j$ in the case when the spectral widths of the photons are narrow and the global time reference is available. This is not surprising, since in this case the duration time of the detection windows is the shortest one (which is obvious since the maximal security distance in this situation is longer than for any other case). In the situation described above, the influence of the timing jitter of the detectors on the key generation rate is visible even for $\tau_j=50\,\mathrm{ps}$, while in all of the other situations the differences begin to appear around $\tau_j=200\,\mathrm{ps}$.

The plots shown in \figref{fig:Qjitter2} were obtained for $\rho=0.99$. It is a reasonable choice, since the influence of the non-zero timing jitter on the results of our QKD security analysis is the strongest for $|\rho|\rightarrow 1$. Thus, one can be certain that if a given value of $\tau_j$ can be neglected during the QKD security analysis when $|\rho|\approx 1$, it can be also neglected for other values of the spectral correlation coefficient.

\section{Correlation and temporal widths}\label{app:C}

\begin{figure}[t]			
	\centering
	\begin{tabular}{c}
		\subfigure[$L=1$ m]{\includegraphics[width=0.32\columnwidth]{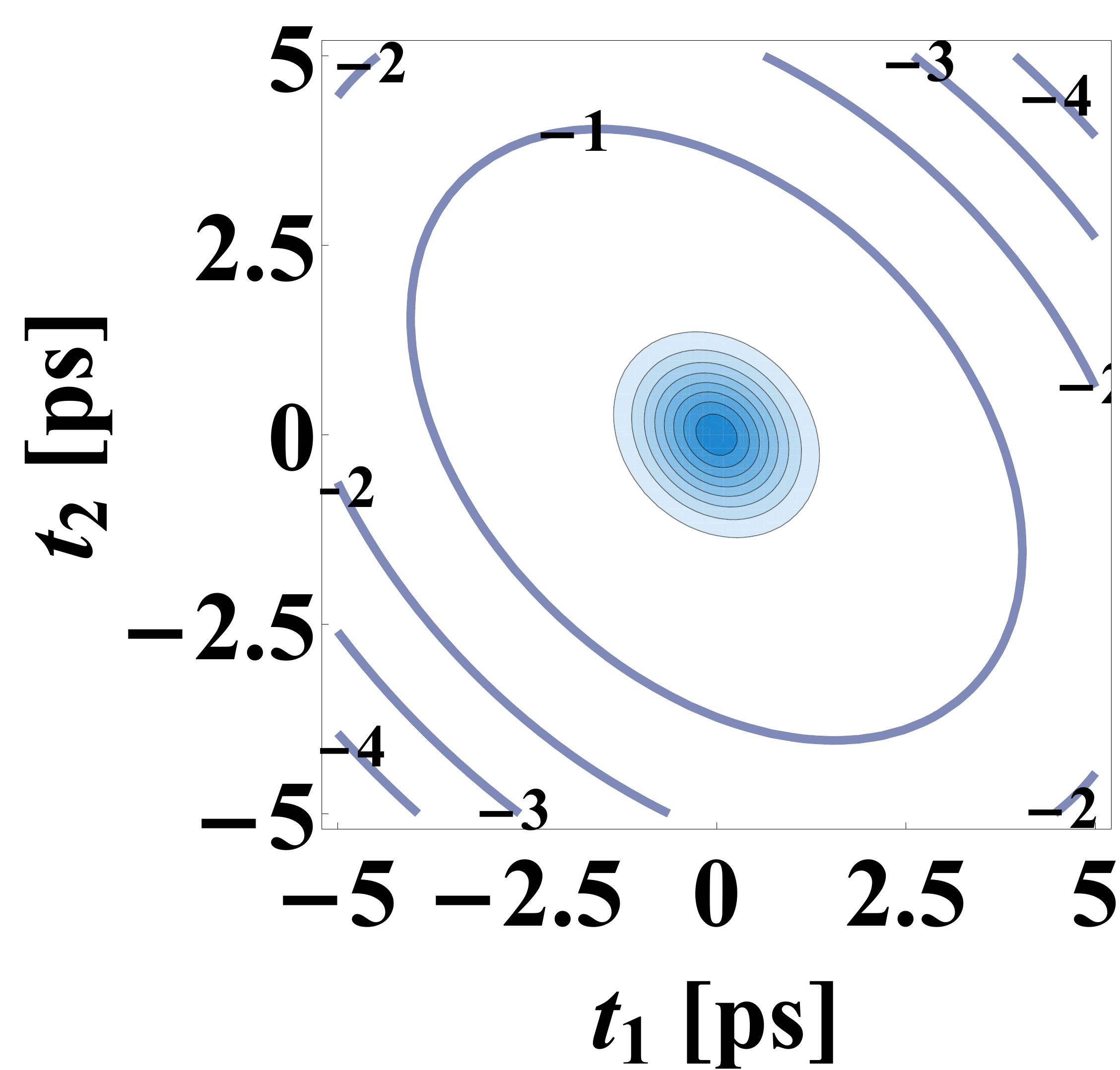}}
		\subfigure[$L=41$ m]{\includegraphics[width=0.32\columnwidth]{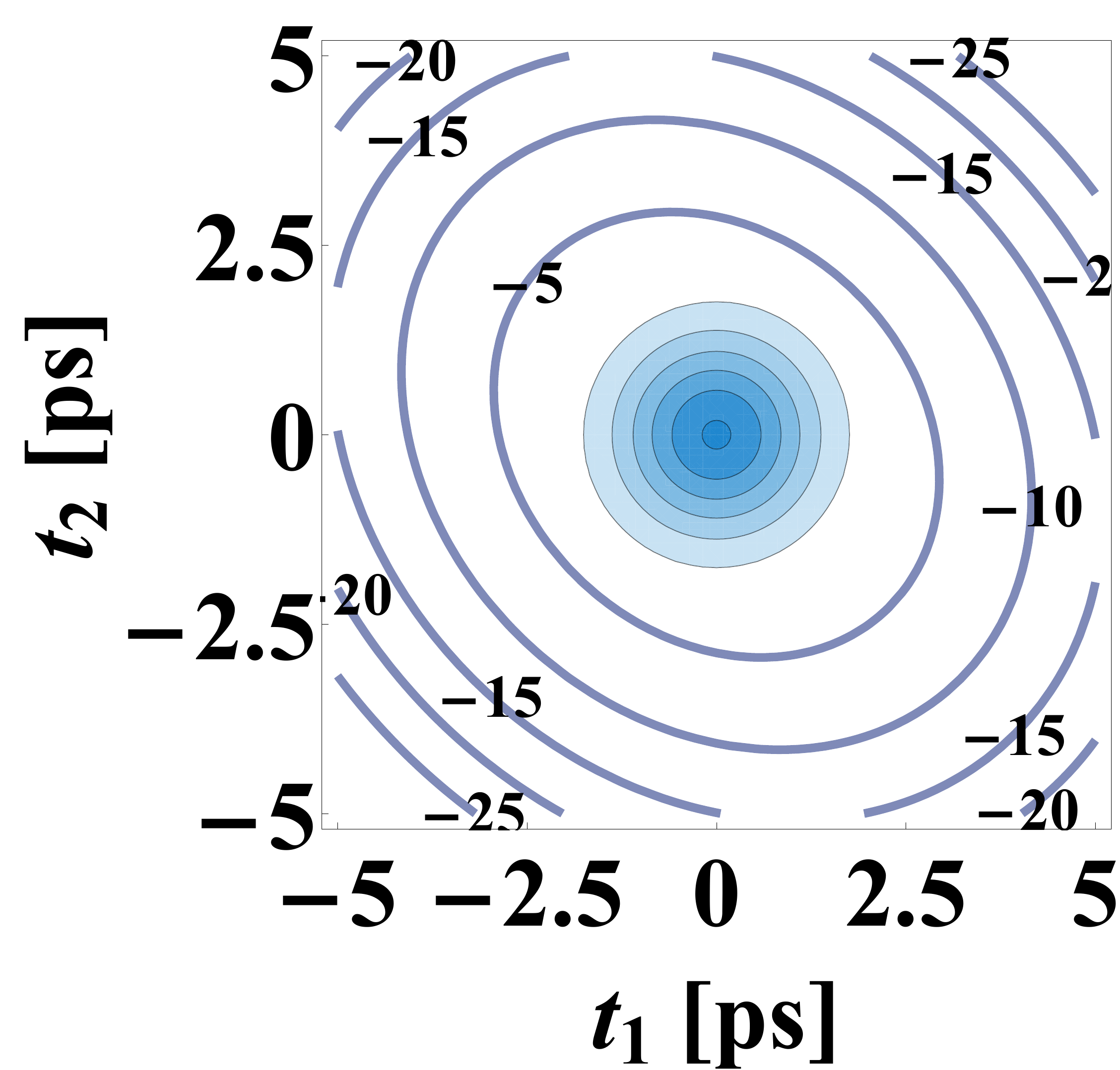}}
		\subfigure[$L=82$ m]{\includegraphics[width=0.32\columnwidth]{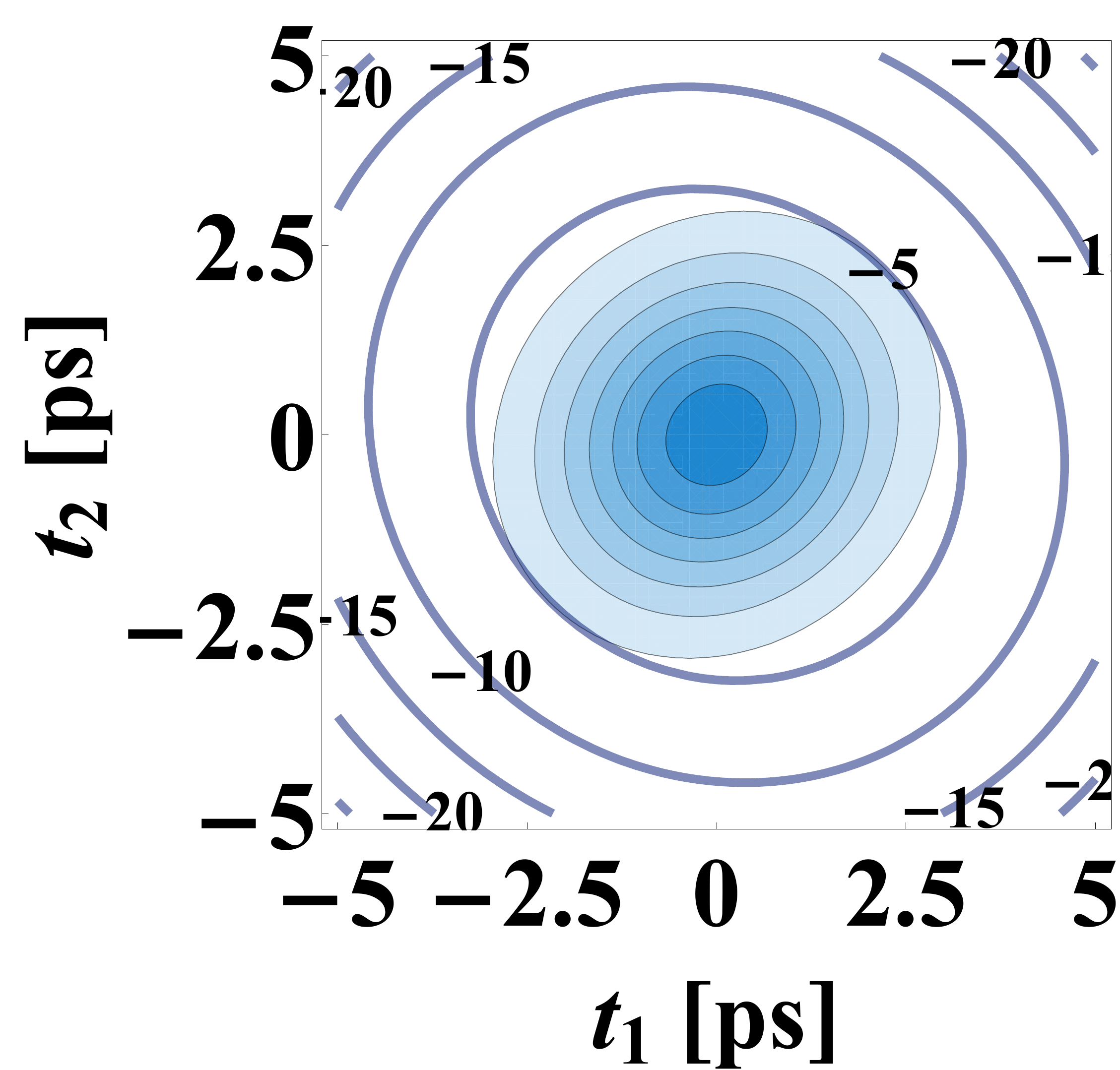}}\\
		\subfigure[]{\includegraphics[width=0.32\columnwidth]{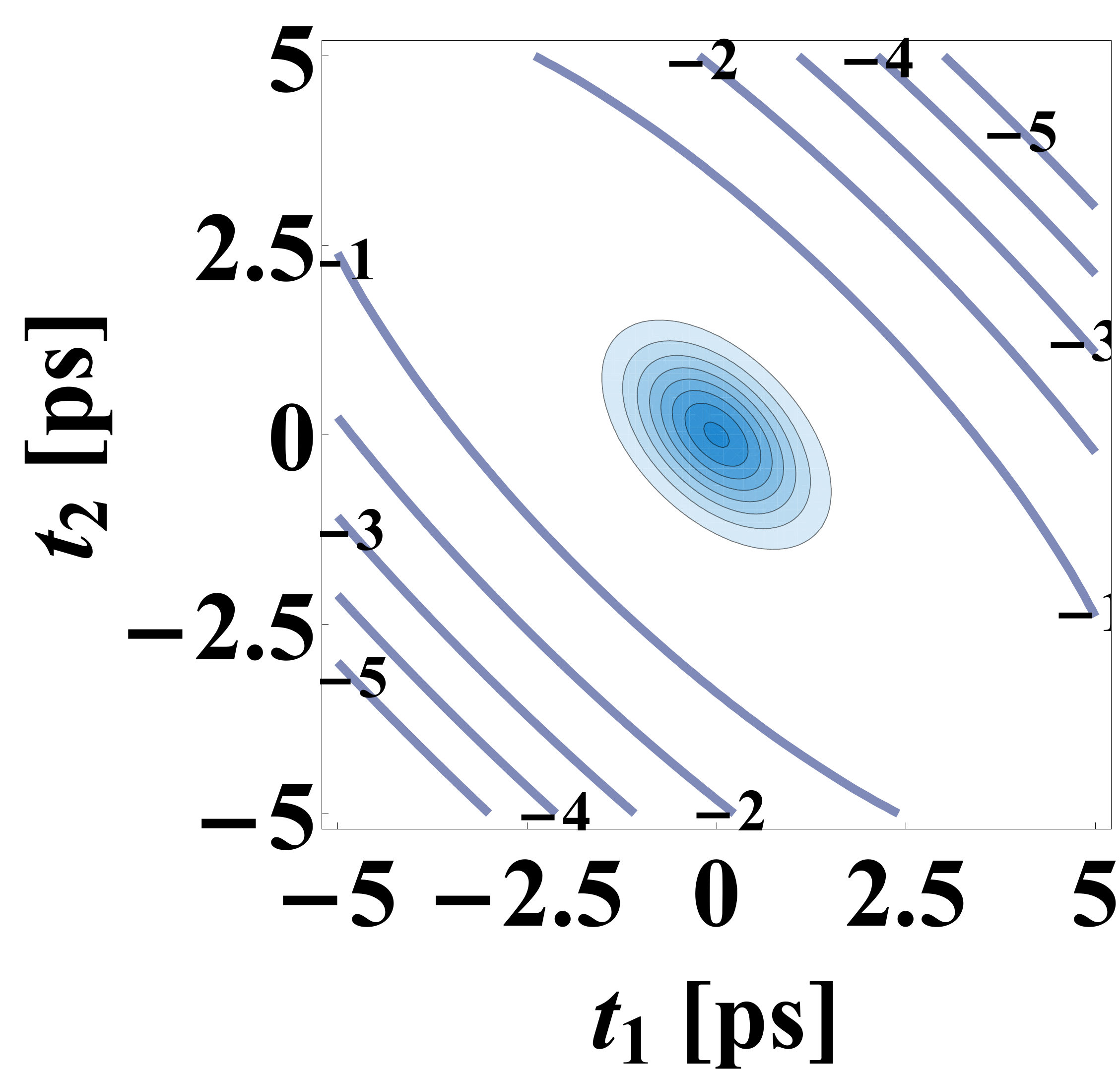}}
		\subfigure[]{\includegraphics[width=0.32\columnwidth]{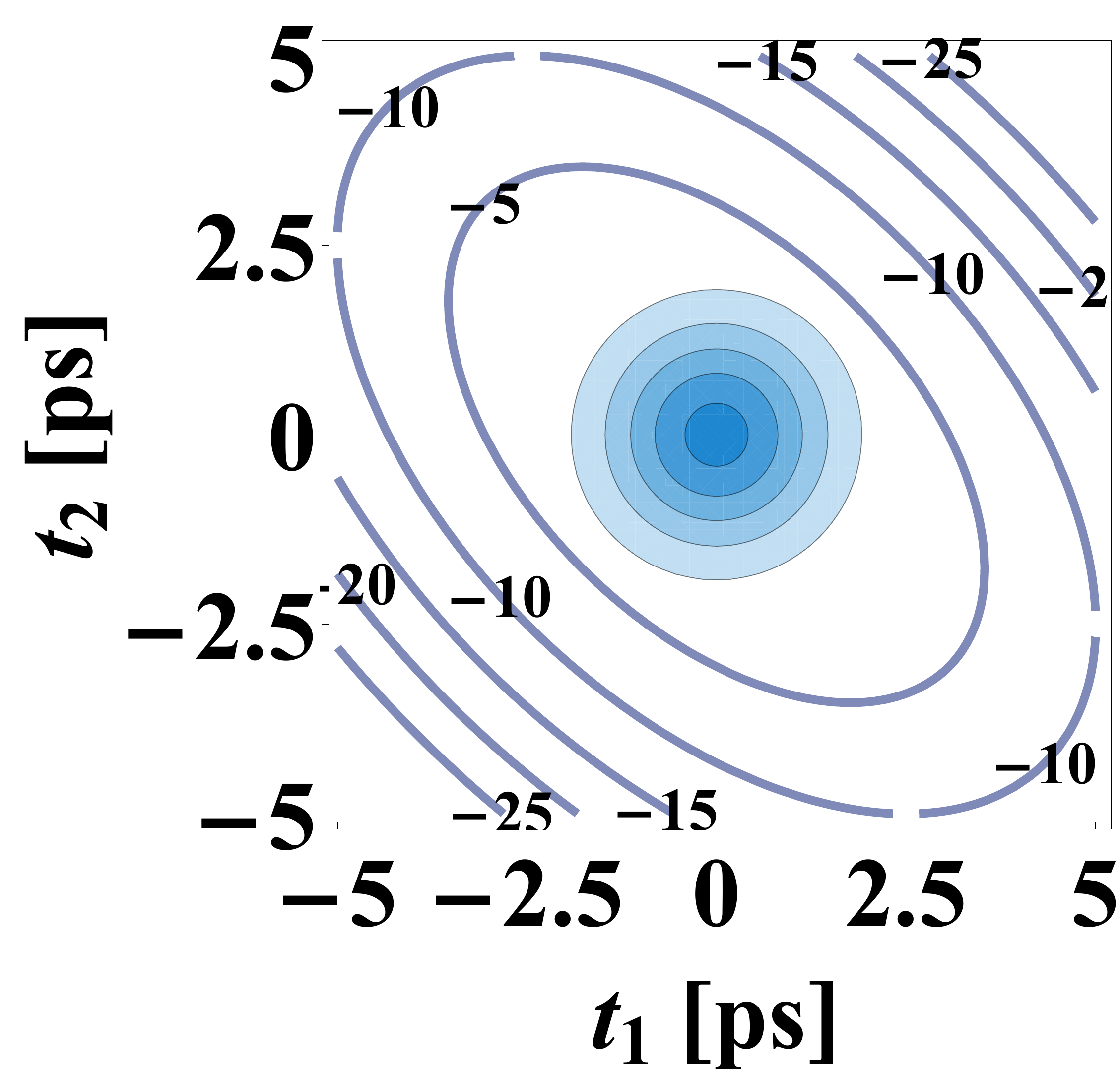}}
		\subfigure[]{\includegraphics[width=0.32\columnwidth]{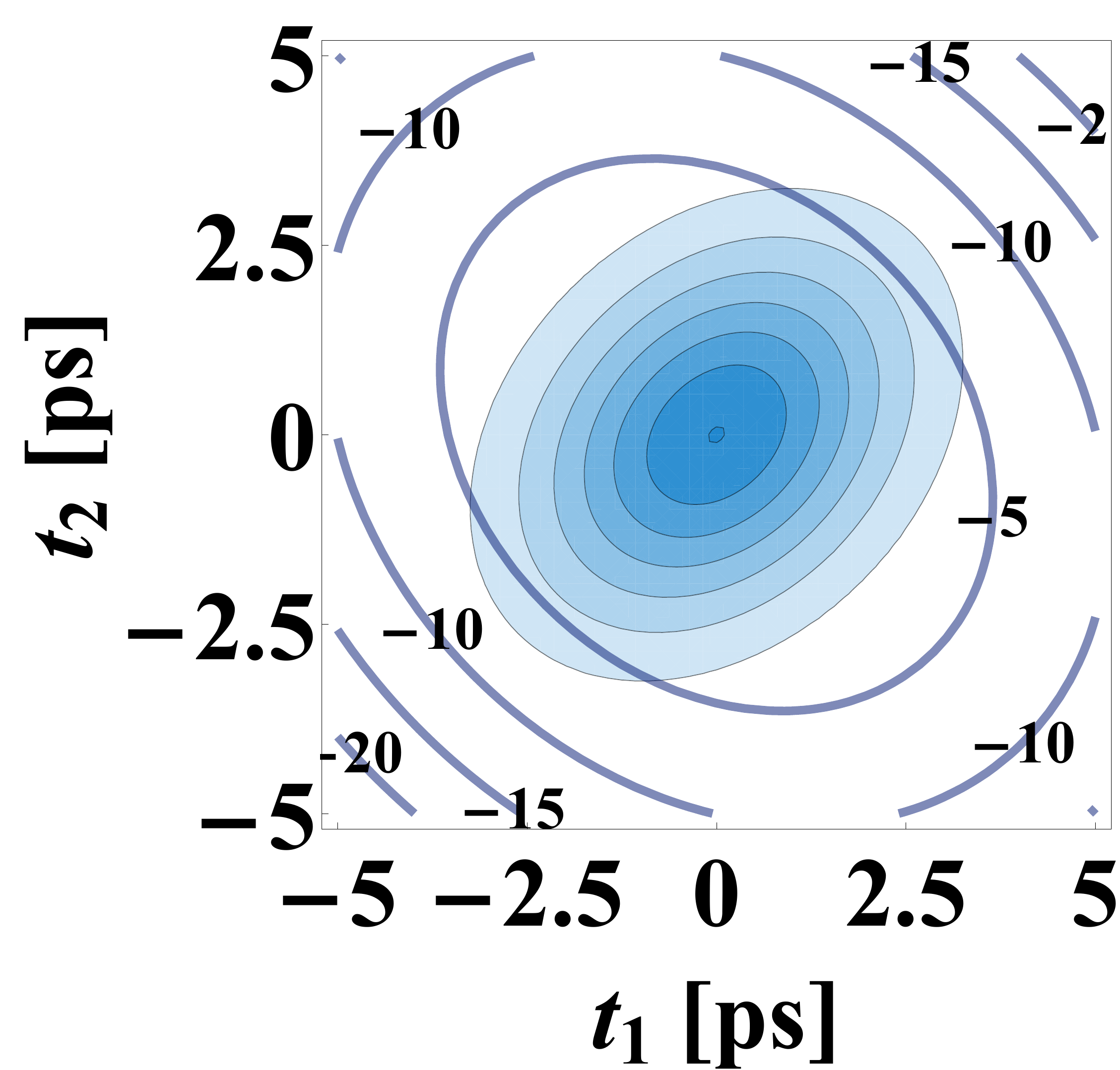}}\\
		\subfigure[]{\includegraphics[width=0.32\columnwidth]{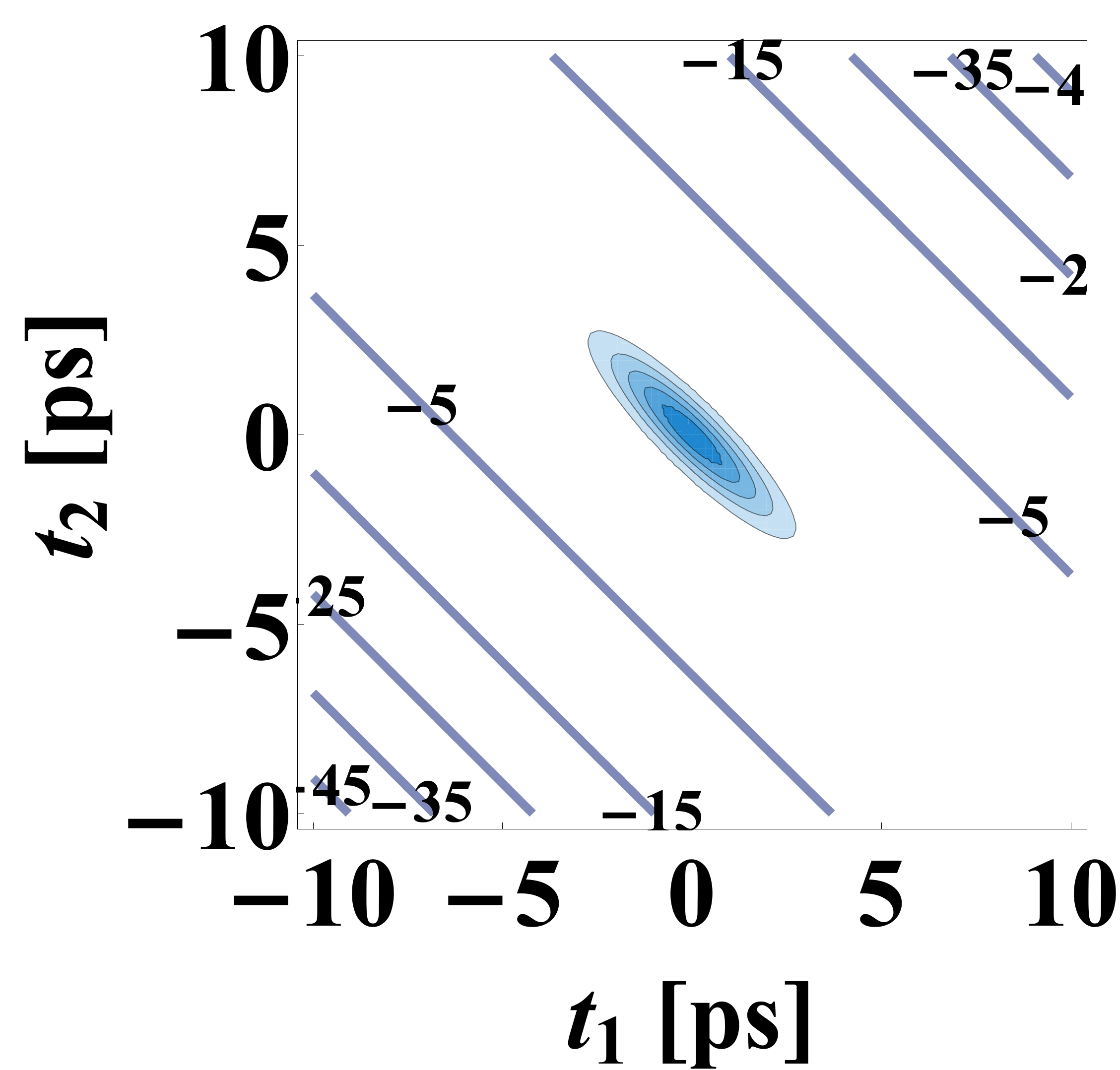}}
		\subfigure[]{\includegraphics[width=0.32\columnwidth]{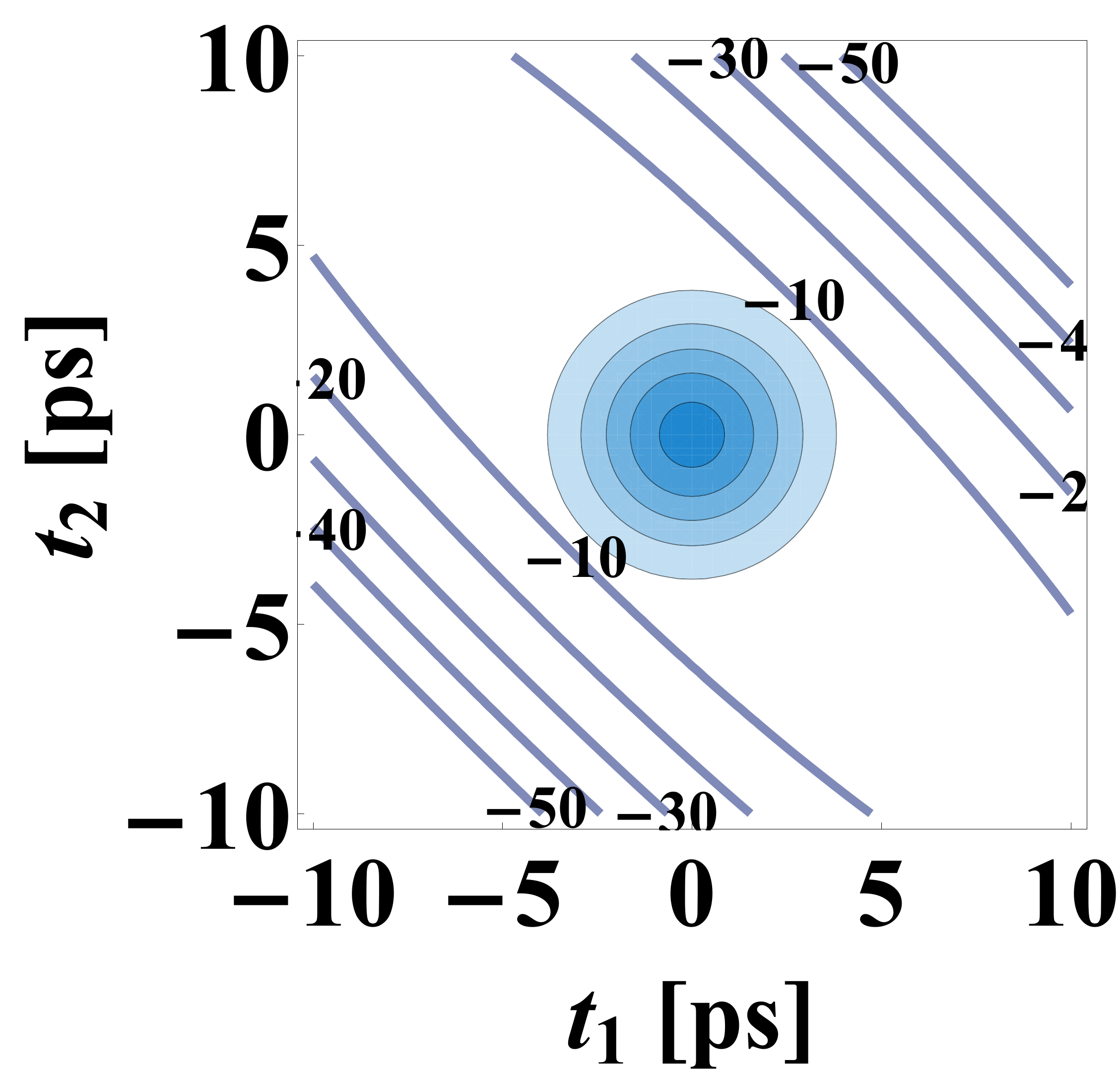}}
		\subfigure[]{\includegraphics[width=0.32\columnwidth]{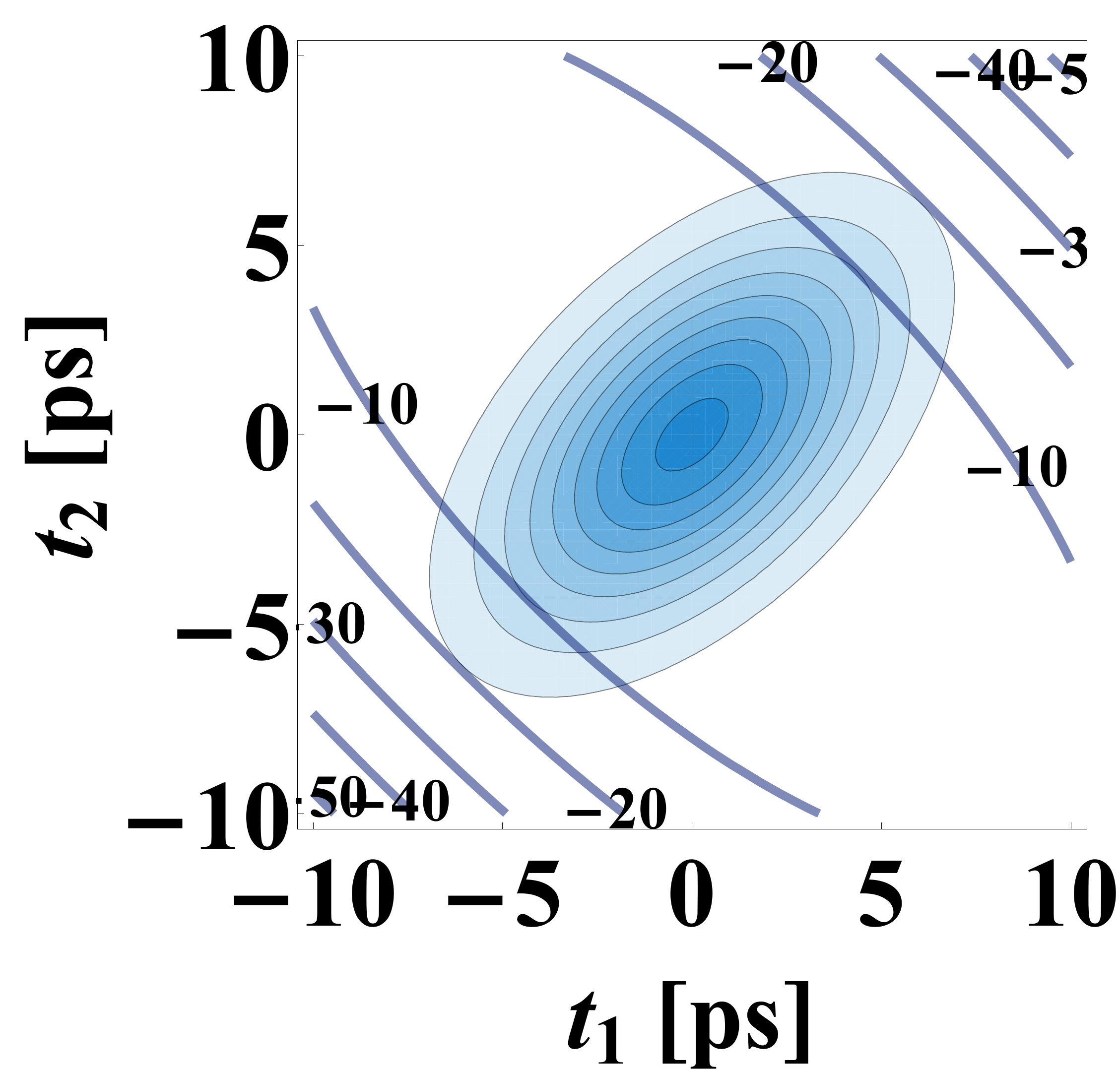}}\\
	\end{tabular}
	\caption{ The amplitude and phase of the biphoton wave function, $\psi_L(t_1,t_2)$. In the columns  the  propagation distances are: $1$ m (a, d, g), $41$ m (b, e, h) and $82$ m (c, f, i). In the rows the initial spectral correlation is taken to be: $\rho=0.2$ (a, b, c), $\rho=0.5$ (d, e, f) and $\rho=0.9$ (g, h, i).}
	\label{fig:psi-L-rho}
\end{figure}

An example of the amplitude and phase of a biphoton wave function for different propagation distance is presented in \figref{fig:psi-L-rho}. The distances,  $L=1$ m, $41$ m and $82$ m, are chosen to demonstrate how the value of temporal correlation changes from negative to positive during the propagation.  

The type of spectral correlation has strong influence on the temporal width of the SPDC photons after the propagation through telecommunication fibers of length $L$. While for the cases when the global time reference is distributed to the owner of a given detection system the temporal widths $\tau_1$ and $\tau_{1h}$ depend on the value of $\rho^2$ (see respectively Eq.~8 and 10). In the opposite situation the temporal width $\tau_{1h,\Delta t}$ depends on $\rho$. Therefore, $\tau_{1h,\Delta t}$ takes different values for different types of spectral correlation, as can be seen in Fig. S3 (c). In this case for short propagation distances the temporal width of the heralded photon arriving at the detection system is the smallest for strong negative spectral correlation between this photon and the heralding photon. On the other hand, for long propagation distances, $\tau_{1h,\Delta t}$ takes the smallest value for strong positive spectral correlation. This behavior of $\tau_{1h,\Delta t}$ is understandable, since for small $L$ positive spectral correlation correspond to the negative temporal correlation, while for large $L$ it is the opposite.
\begin{figure}[t]
	\centering
	\begin{tabular}{c}
		\subfigure[]{\includegraphics[width=0.49\columnwidth]{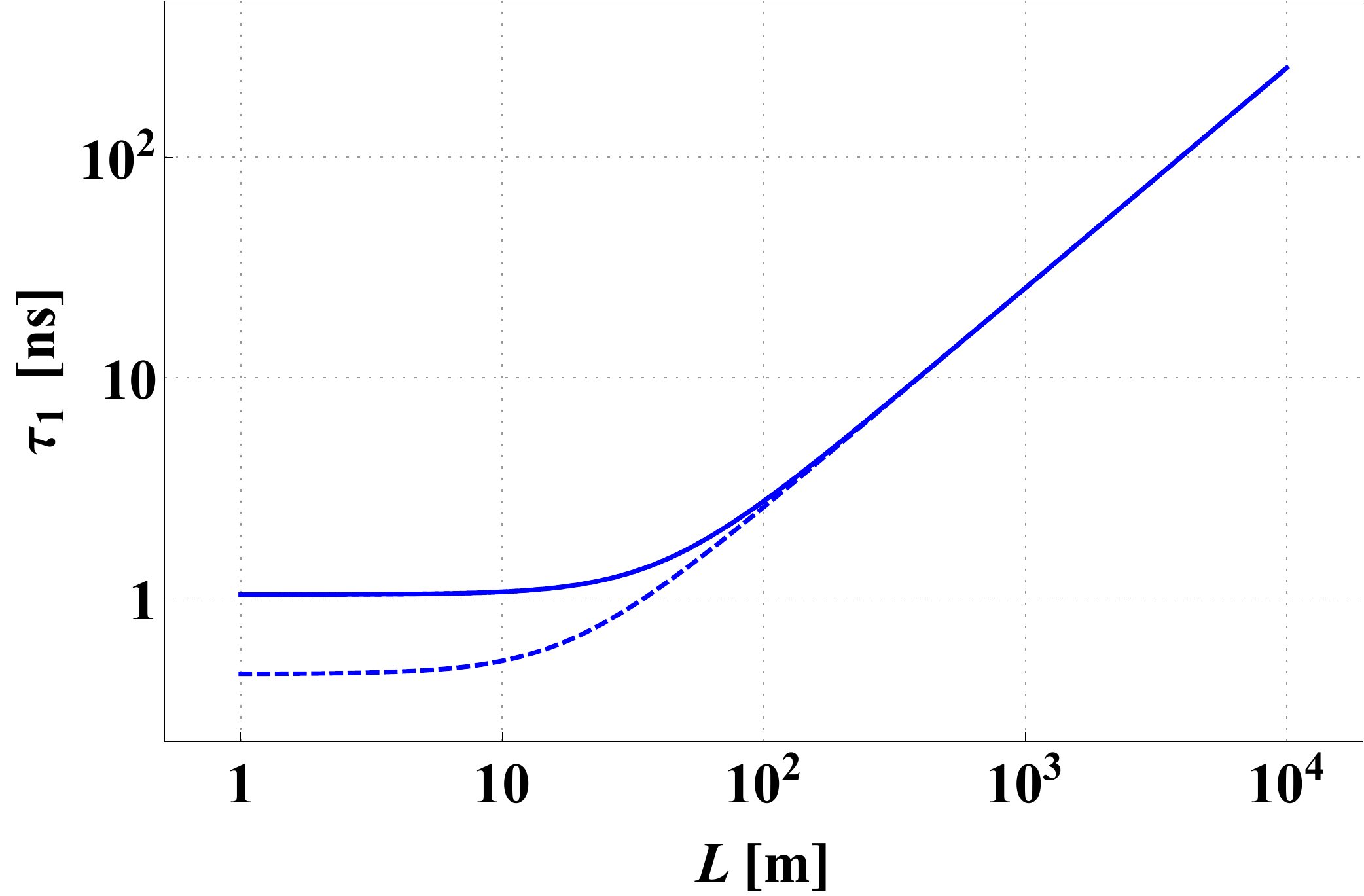}}
		\subfigure[]{\includegraphics[width=0.49\columnwidth]{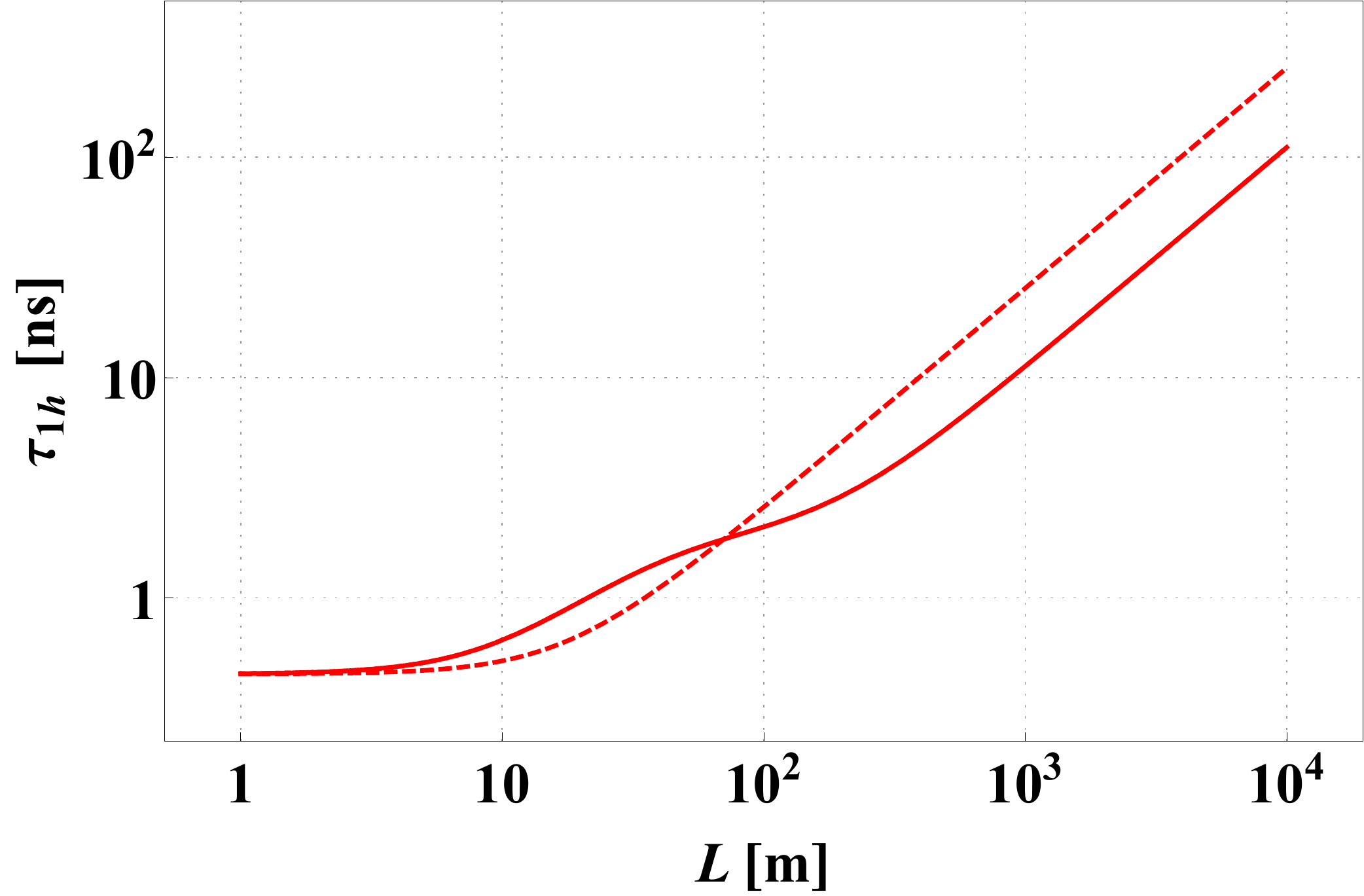}}\\
		\subfigure[]{\includegraphics[width=0.49\columnwidth]{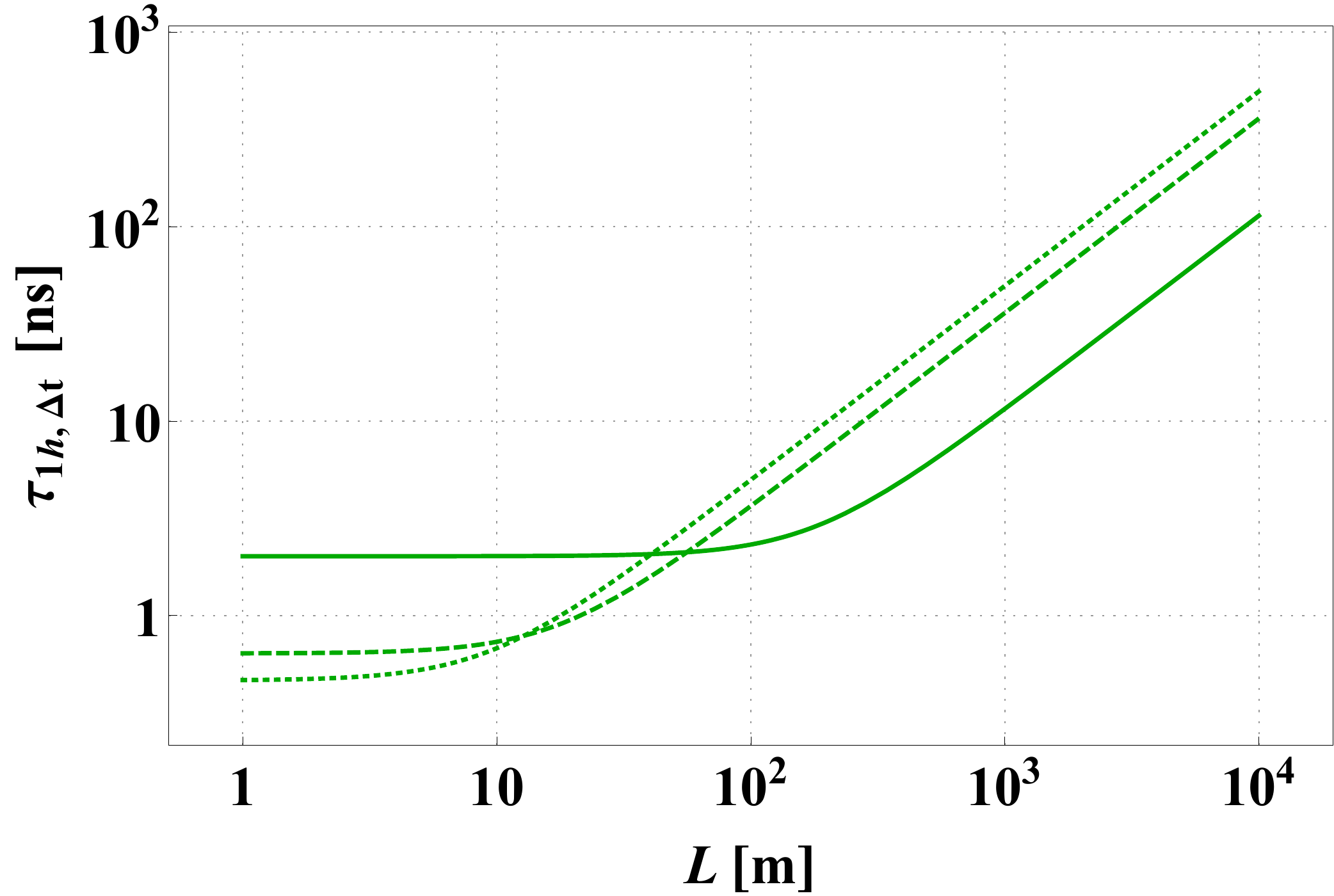}}
	\end{tabular}	
	\caption{The temporal widths of one of the photons created in the SPDC process after both of these photons propagate through telecommunication fibers of length $L$ in the three scenarios described in the Sec. 4 of our article.: no-heralding scenario with available global time reference (panel (a)), heralding scenario with available global time reference (panel (b)) and heralding scenario with unavailable global time reference (panel (c)). The solid, dashed and dotted lines are drawn for $\rho=0.9$, $\rho=0$ and $\rho=-0.9$ respectively. On panels (a) and (b) the dotted lines are invisible, because they are the same as solid lines.  }
	\label{fig:det-times}
\end{figure}
\\ \indent The situation is different for the temporal widths $\tau_1$ and $\tau_{1h}$, plotted in Fig. S3 (a) and (b), respectively. Both of these two functions depend only on the strength of the spectral correlation coefficient, not on its sign. In the context of application in quantum communication protocols, long propagation distances ($L>10^3$ m) are much more interesting to us, than the short ones. Therefore, it is worth noting here that for $L>10^3$ m $\tau_1$ is generally independent of $\rho$, while $\tau_{1h}$ decreases when the absolute value of $\rho$ increases.\\
\indent All of the above conclusions, stemming from the analysis of Fig. S3, are consistent with the results of the security analysis of the QKD scheme,  which were presented in Fig. 5 (a).
%In the case called the no-heralding scenario, in which the detection time of photon number 2 is not available, the temporal width of photon number 1, $\tau_1(\sigma_1)$, depends on $\rho^2$, see Eq.~8. Next, if the detection time of photon number 2 is known to be $t_2$, a temporal width of photon number 1, $\tau_{1h}(\sigma_1,\sigma_2)$, has also $\rho^2$ dependence, see Eq.~10.

%Figure \ref{fig:det-times} depicts the temporal widths as a function of propagation distance. In the context of security analysis we  focus on the long distances ($L>10^3$ m). Once can see in panel (a) that the spectral correlation has no impact on the temporal width, $\tau_1(\sigma_1)$, in the no-heralding scenario. On the other hand, when the detection time of heralding photon is known, the temporal width, $\tau_{1h}(\sigma_1,\sigma_2)$ is narrower, when the pair is correlated as can be seen in panel (b). Finally, in panel (c) one can see that the temporal width, $\tau_{1h,\Delta t}(\sigma_1,\sigma_2)$, is the narrowest for the positively  correlated photons. Note that the negative correlation broadens the temporal width compared to the case of no correlation.

\end{document}